\documentclass[useAMS,usenatbib]{mn2e}

\usepackage[pdftex]{graphicx}

\newcommand\diamondrule{\line{$\m@th
   \leaders\hrule\hfill\rlap{$\m@th\bracerd\braceld$}
   \braceru\bracelu\leaders\hrule\hfill$}}

\newcommand\smallneg{\kern-.0800em}
\newcommand\negskip{\kern-.5em}

\newcommand\lsim{\rlap{\raise.4ex\hbox{$<$}}\lower.55ex\hbox{$\sim$}\,}
\newcommand\gsim{\rlap{\raise.4ex\hbox{$>$}}\lower.55ex\hbox{$\sim$}\,}
\newcommand\implies{{\bf=\kern-0.45em>}}

\newcommand\unit{\,\rm}

\newcommand\kms{\rm\, km\cdot s^{-1}}
\newcommand\Kkms{\rm\, K\cdot\kms}
\newcommand\nkms{\, km\cdot s^{-1}}
\newcommand\um{\,\mu m}

\newcommand\Jtwo{\rm J=2\rightarrow 1}

\newcommand\Jone{\rm J=1\rightarrow 0}

\newcommand\NH{\rm N(H_2)}

\newcommand\cKkms{\unit cm^{-2}\cdot (K\cdot km\cdot s^{-1})^{-1}} 
\newcommand\mfif{M$\,$51}
\newcommand\meig{M$\,$83}
\newcommand\Nxh{N_x(H_2)}
\newcommand\Xf{X_F}
\newcommand\Xfm{X_{FM}}
\newcommand\Xtw{\, X_{20}}
\newcommand\rchi{\chi_\nu^2}
\newcommand\mjysr{$\, MJy\cdot sr^{-1}$}
\newcommand\mjsr{\, MJy\cdot sr^{-1}}
\newcommand\es{erg\cdot s^{-1}}
\newcommand\esh{erg\cdot s^{-1}\cdot Hz^{-1}}

\newcommand\escs{erg\cdot s^{-1}\cdot cm^{-2}\cdot sr^{-1}}
\newcommand\Td{T_d}
\newcommand\Nd{N_d(H)}
\newcommand\Inu{I_\nu}

\newcommand\Bnu{B_\nu}

\newcommand\knu{\kappa_\nu}

\newcommand\knup{\kappa_\nu(1.1\,mm)}
\newcommand\kno{\kappa_{\nu_o}}
\newcommand\cmg{\, cm^2\cdot g^{-1}}

\newcommand\taun{\tau_{\nu d}/N(H)}
\newcommand\mspc{\, M_\odot\cdot pc^{-2}}

\newcommand\msyrkp{\, M_\odot\cdot yr^{-1}\cdot kpc^{-2}}
\newcommand\Sg{\Sigma_{gas}}
\newcommand\SH{\Sigma_{H2}}
\newcommand\Sh{\Sigma_{HI}}
\newcommand\Sfr{\Sigma_{SFR}}

\newcommand\SFR{\hbox{\.M}_\ast}
\newcommand\Sd{\Sigma_d(gas)}

\title[{\it AzTEC\/} \mfif/\meig\ Observations]
{Continuum Observations of \mfif\ and \meig\ at 1.1$\,$mm with\\ {\it AzTEC\/}}
\author[W. F. Wall et al.]
  {W.~F.~Wall,$^1$
  I.~Puerari,$^1$ R.~Tilanus,$^2$ F.~P.~Israel$^2$ J.~E.~Austermann,$^3$ I.~Aretxaga,$^1$ 
  \newauthor 
   G.~Wilson,$^4$ M.~Yun,$^4$ K.~S.~Scott,$^5$ T.~A.~Perera,$^6$ C.~M.~Roberts$^4$ 
  \newauthor
   and D.~H.~Hughes,$^1$ 
 \\
  $^1$Instituto Nacional de Astrof\'{\i}sica, \'Optica, y Electr\'onica, Apdo. 51 y 216, Puebla, Puebla, M\'exico\\
  $^2$Leiden Observatory, Leiden University, NL 2300 RA Leiden, The Netherlands\\
  $^3$NIST Quantum Devices Group, 325 Broadway Mailcode 817.03, Boulder, CO, USA 80305\\
  $^4$Department of Astronomy, University of Massachusetts, Amherst, MA 01003, USA\\
  $^5$North American ALMA Science Center, NRAO, 520 Edgemont Rd, Charlottesville, VA, USA 22903\\
  $^6$CNS C 007C, Illinois Wesleyan University, Bloomington, IL, USA 61702-2900
}
\date{Accepted 2016 March 21. Received 2016 March 4; in original form 2015 November 23}

\pagerange{\pageref{firstpage}--\pageref{lastpage}} 

\def\LaTeX{L\kern-.36em\raise.3ex\hbox{a}\kern-.15em
    T\kern-.1667em\lower.7ex\hbox{E}\kern-.125emX}

\begin{document}

\label{firstpage}

\maketitle

\begin{abstract}

We observed the spiral galaxies \mfif\ and \meig\ at 20$"$ spatial resolution
with the bolometer array {\it AzTEC\/} on the {\it JCMT\/} in the 1.1$\,$mm continuum,
recovering the extended emission out to galactocentric radii of more
than 12$\,$kpc in both galaxies. The 1.1$\,$mm-continuum fluxes are
$5.6\pm 0.7$ and $9.9\pm 1.4\, Jy$, with associated gas masses estimated at
$9.4\times 10^9\, M_\odot$  and $7.2\times 10^9\, M_\odot$ for \mfif\ and \meig, respectively.
In the interarm regions of both galaxies the $\NH/I(CO)$  (or X-factor)
ratios exceed those in the arms by factors of $\sim 1.5$-2.  In the inner
disks of both galaxies, the X-factor is about $1\times 10^{20}\cKkms$.
In the outer parts, the CO-dark molecular gas becomes more important.

While the spiral density wave in \mfif\ appears to influence the
interstellar medium and stars in a similar way, the bar potential in \meig\ 
influences the interstellar medium and the stars differently. We
confirm the result of \citet{Foyle10} that the arms merely heighten the
star formation rate and the gas surface density in the same proportion.  Our maps
reveal a threshold gas surface density for an SFR increase by two
or more orders of magnitude. In both galaxy centers, the molecular gas
depletion time is about 1$\,$Gyr climbing to 10-20$\,$Gyr at radii of 
6-8$\,$kpc. This is consistent with an inside-out depletion of the molecular
gas in the disks of spiral galaxies.

\end{abstract}

\begin{keywords}
 galaxies: ISM -- galaxies: spiral -- galaxies: individual: \mfif, \meig 
\end{keywords}

\section{Introduction}

Central to a complete picture of galaxy evolution is the distribution of the interstellar matter (ISM) within each galaxy and how that ISM forms stars.
Given that the ISM mass on galactic scales is dominated by molecular and atomic gas, observing the tracers of these gas components is necessary for 
measuring the ISM distribution within the disks of spiral galaxies.  Accordingly, observations of the HI 21-cm line and the CO~$\Jone$ 2.6-mm line are 
often used as tracers of the atomic and molecular gas, respectively \citep[see, e.g.,][]{Walter08, Regan06, Regan01, Nish01}.  While conversion of
the velocity-integrated brightness temperature of the
HI line, or $I(HI)$, to atomic gas column density, $N(HI)$, is usually straightforward \citep[though not always, e.g.,][]{Parade11a}, the conversion of 
I(CO) to molecular gas column density $N(H_2)$ is not quite so certain \citep[e.g.,][]{Papa12, Nar11, Nar12,  Shetty11, Shetty11a, Maloney88, Rickard85, 
Israel88, W93, Regan00, Pag01, Sodroski95, Dahmen97, Dahmen98, W07}, especially given that the CO~$\Jone$ line is known to be optically thick 
\citep[e.g., see][]{Evans80, Kutner84, Evans99}.   Recent observations \citep{Parade11a} suggest that the $N(H_2)/I(CO)$ conversion factor,
or X-factor, $\Xf$, is roughly constant within the disk of our Galaxy, with $\Xf = (2.5\pm 0.1)\times 10^{20}\, H_2\cdot cm^{-2}\cdot (K\cdot\nkms)^{-1}$, 
to be abbreviated as $\Xf = 2.5\pm 0.1\,\Xtw$; this or a similar value of $\Xf$ is often called the ``standard'' value.   This uniform X-factor value 
for our Galaxy's disk now applies
to the disks of external galaxies, where $\Xf\sim 2\,\Xtw$ is inferred and, on average, is {\it radially\/} non-varying from the 
inner disk to a galactocentric radius of 
$\sim 1\, R_{25}$ \citep{Sandstrom13}.
The evidence for CO-dark gas, both theoretically and observationally,
is a further complication \citep[see, for example][]{Parade11a, Clark12, Langer14, Saintonge12, Smith14, Roman10}.   It is thus advantageous
to employ tracers other than CO~$\Jone$ as independent checks on ISM surface density variations to test recent physical models of $\Xf$ 
\citep[e.g.,][]{Papa12, Nar11, Nar12,  Shetty11, Shetty11a}

Observed gas column and surface densities can then provide insights into large-scale star formation in galaxies.   The Schmidt-Kennicutt (S-K) law,
for example, states that star formation rate surface density, $\Sfr$, is related to the gas surface density, $\Sg$, by
$\Sfr\propto\Sg^\alpha$ with $\alpha\sim 1.0$ to $\sim 5.0$ \citep[e.g.,][]{Schmidt59, Kennicutt89, Kennicutt98, Bigiel08, Heiderman10, Leroy13}.
The index $\alpha =1.0$ is appealing because the gas depletion time  ($\propto\Sg/\Sfr$) is constant thoughout the spiral disks.  While there is some 
evidence that $\alpha =1$ on size scales of $\gsim 1\, kpc$ 
\citep[e.g.][]{Bigiel08,Leroy13,Calzetti12}, there is also evidence of non-linear and even non-universal slopes
on such size scales \citep[see, e.g.][]{Santini14, Pan14, Shetty13}, especially on the scales of individual giant molecular clouds (GMCs) 
\citep[][]{Lombardi14, Lada13}.  There is also strong evidence for an inside-out formation of galaxies \citep[e.g.][]{Gonzalez14}, which is at
odds with a constant gas depletion time and, therefore, with having $\alpha = 1$.  
Comparison between those results on the large (i.e., galactic) scales with those on GMC scales is problematic.
\citet{Lada13} suggest that the Schmidt law on the scales of GMCs are fundamentally different from the S-K law apparent on larger 
(i.e. galactic) scales; the latter are not the ``result of an underlying physical law of star formation.'' 

 
Many of the abovementioned results used observations of the optically thick CO~$\Jone$ line and adopted a spatially 
constant $\Xf$.  In contrast, millimetre (mm), submillimetre (submm), and far-infrared (far-IR) continuum observations sample optically 
thin continuum emission from the dust grains that pervade both the atomic 
and molecular gas.  Recently, there have been many papers of the far-IR/submm continuum emission of external galaxies from the {\it Planck} and 
{\it Herschel} missions \citep[e.g.][]{Parade11, Foyle12, Fritz12, Foyle13, Magnelli12, Bourne12, Rowlands12, Boselli12, Smith12,  Cortese14, Cortese12, Boselli10,
Eales10, Gordon10, Roman10, Kirkpatrick14}.
These papers find, for example, that the dust and stellar masses of galaxies are correlated \citep{Bourne12, Cortese12} and that spiral galaxies and
dusty early type galaxies have $\sim 10^6$ to $10^8\, M_\odot$ of dust \citep{Rowlands12}.   
 
A major stumbling block to determining accurate surface densities from dust continuum emission is the unknown mass absorption coefficient, $\knu$, 
at millimetre wavelengths.   Millimetre continuum emission is less temperature sensitive than that 
at submillimetre and far-IR wavelengths; this provides an important constraint on dust mass and sometimes the spectral emissivity index, 
$\beta$, can be constrained as well.  
Observationally, the relevant quantity determined is the dust optical depth to gas 
column density ratio, $\taun$.  \citet{Parade11a} and \citet{Parade11b} have found that $\taun=5.2\times 10^{-26} cm^2$ at 857$\,$GHz (wavelength of 
350$\um$) in the HI gas in the solar neighbourhood and  $\taun=1.1\times 10^{-25} cm^2$ at 250$\um$ 
(corresponding to $\taun=6.0\times 10^{-26} cm^2$ at 857$\,$GHz for
$\beta=1.8$) in the HI gas in the Taurus molecular complex.  Given that $\beta=1.8$ applies to
the dust in our Galaxy \citep[see, e.g.,][]{Parade11a} and that the dust to {\it hydrogen\/} gas mass is about 0.01, those observed $\taun$ correspond to
$\knup\simeq 0.4$ to 0.5$\cmg$ in the dust associated with HI.  The dust associated with H$_2$, however, has $\taun$ double that in HI
\citep{Parade11b}.  Consequently, estimating $\taun$ and $\knu$ from comparing the observed dust continuum emission against the HI gas emission alone, 
while useful, must be viewed with caution.   The various estimates of $\knu$ suggest that determining the exact {\it total\/} mass of dust within a galaxy is 
uncertain by a factor of a few.   

In spite of the uncertain dust mass absorption coefficient, dust continuum emission can provide estimates of $\Xf$ in our Galaxy as well as in external 
galaxies \citep[see, e.g.,][]{Sandstrom13, Roman10, Eales10, Reach98, Parade11a, Israel97, Israel97a}.  Such observations have shown that while $\Xf$ 
can be more or less 
spatially constant in some cases, like in the disk of our Galaxy and other external galaxies \citep[see][]{Sandstrom13, Parade11a}, there can be 
regions of ``dark'' gas, H$_2$ with no CO emission, both 
in our Galaxy and other galaxies \citep[e.g., see][]{Parade11a, Baes14, Clark12, Langer14, Pineda14, Roman10, Saintonge12, Smith14}.  
Therefore, observations of dust continuum emission provide a vital check on results inferred from 
CO~$\Jone$ observations.

\subsection{The Current Work}
Even with the many recent advances mentioned above, there are many questions left unanswered.  For example, do the inferred 
X-factor values \citep[i.e.][]{Parade11a, Sandstrom13} apply to the outer disks of all galaxies?  Also, are there systematic differences of the 
X-factor between arm and inter-arm regions?  Can previous methods of observationally inferring the dust mass absorption coefficient 
at millimetre wavelengths be refined?   How do the answers to those questions influence the specific form of the observed
S-K law in a given galaxy?  

To address these questions and to 
better understand the gas and dust in spiral galaxies and their relationship to star formation, we observed the grand-design, 
face-on spiral galaxies \mfif\ and \meig\
with the bolometer-array camera, {\it AzTEC\/} (Aztronomical Thermal Emission Camera), mounted on the 15-m JCMT in Hawaii at a 
wavelength of 1.1$\,$mm. 
Both of these galaxies are nearby with distances of less than 10$\,$Mpc (see Table~\ref{tab1} for details) and, hence, the spiral 
arms in both galaxies are 
resolved across the arms in the {\it JCMT/AzTEC\/} observations, which have a spatial resolution of 20$''$.  Both galaxies have 
been studied extensively at numerous
wavelengths 
\citep[e.g.,][]{Rots90, Helfer03, Kennicutt03, Dale09, Tilanus93, Crosth02, Blasco10, Meidt13, Hughes13, Schinnerer13, Colombo14}.   

Recent {\it HERSCHEL\/} observations of 
\mfif\ and \meig\ at 70, 160, 250, 350, and 500$\um$ have provided maps of the dust temperature, surface density, and even 
spectral emissivity index, $\beta$ 
\citep{Bendo12, Foyle12, Foyle13, Cooper12}, but with spatial coverage that is slightly more limited than those of the  {\it AzTEC\/} 
1.1$\,$mm continuum 
maps presented here; the {\it AzTEC\/} images cover a few more kiloparsecs at the adopted distances given in Table~\ref{tab1}.
As a result, these  {\it AzTEC\/} 1.1$\,$mm continuum images extend both the spatial and the wavelength coverage of the dust 
emission in both \mfif\ and \meig.   

This greater spatial coverage placed restrictions on the other wavelengths available for comparison with the {\it AzTEC\/} 
1.1$\,$mm data; there were no  250, 350, and 500$\um$ data towards the outer disks of \mfif\ and \meig.  There are, however,
{\it Spitzer\/} 160$\um$ data covering both of these galaxies.  This wavelength is the longest of the {\it Spitzer\/} data and,
along with the {\it AzTEC\/} 1.1$\,$mm data, is the one most likely associated with the dust component(s) that dominate the mass of 
the dust.   

In addition, the surface densities in the current work were estimated
from the observational data.  Specifically, $\taun$ at $\lambda =1\, mm$ was inferred by comparison with the HI column densities
and removing upper outliers, because these upper outliers were assumed to represent positions with CO-dark gas.  This approach
has the advantage that final gas masses inferred were not dependent on dust models. 



%


\begin{table}
 \caption{Adopted Parameter Values}
 \label{tab1}
 \begin{tabular}{@{}lll}
  \hline
  Parameter & \mfif & \meig \\
  \hline
  Centre Position (2000.0) & $13^{\rmn{h}}29^{\rmn{m}}52\fs71\, ^a$ & \phantom{$-$}$13^{\rmn{h}}37^{\rmn{m}}00\fs8\, ^b$ \\
				      & $47\degr 11\rlap{$\arcmin$}^{\phantom{\rmn{m}}} 42\farcs 80\, ^a$ & $-29\degr 51\rlap{$\arcmin$}^{\phantom{\rmn{m}}} 59\arcsec\, ^b$ \\
  Distance (Mpc)		      & 8.4$\, ^c$ & 4.5$\, ^d$ \\
  Position Angle		      & $170\degr\, ^e$ & $225\degr\, ^f$ \\
  Inclination		      & $20\degr\, ^e$ & $24\degr\, ^f$ \\ 
  \hline
 \end{tabular}

 \medskip
$^a$ \citet{Hagiwara01}.

$^b$ \citet{Miller09}.

$^c$ See \citet{Shetty07} and references therein. 

$^d$ \citet{Karach02}.

$^e$ \citet{Tully74}.

$^f$ \citet{Talbot79}.

\end{table}

\section{Observations}

Both \mfif\ and \meig\ were observed during the nights of 2005 December 6-12 and 2006 January 12-20 on the 
James Clerk Maxwell Telescope (JCMT) with {\it AzTEC\/} (Aztronomical Thermal Emission Camera), which is a bolometer 
array for observing continuum emission at 1.1$\,$mm \citep{Wilson08}.   {\it AzTEC\/} has 144 silicon nitride micromesh
detectors arranged hexagonally in six ``hextants''.  The {\it AzTEC\/} foot-print on the sky while mounted on the JCMT was 
$\sim$280$''$ wide, where
each detector had an 17-18$''$ almost circular beam.  During the observations, 107 of the 144 {\it AzTEC\/} detectors were 
fully functional, the faulty detectors found mostly in Hextants~1 and 2 \citep[see Figure~11 of][]{Wilson08}.  The 
images of \mfif\ and \meig\ were created by raster scanning, so the missing detectors did not affect the coverage 
of the final images, only their sensitivities.

The fields for both galaxies were originally chosen to be centred on each and were $14'\times 14'$ in size.   This field size 
is large enough to accommodate each galaxy with a $\sim 10'$ diameter, and one-half {\it AzTEC\/} foot-print on each side.
The actual observations, however, covered nearly $25'\times 25'$ for each galaxy. 
This allows the {\it AzTEC\/} routines to properly remove the large-angular-scale atmospheric emission from the 
\mfif\ and \meig\ images.  
Both \mfif\ and \meig\ were raster scanned for many cycles; the total integration time was about 12 hours for \mfif\ and about 
14.5 hours for \meig. {\it Given the long-term stability of the detectors, chopping was not necessary.\/}  
During the observations, the zenith optical depth at 225$\,$GHz was between about 0.05 and 0.15.
The source elevations during the observations were between about 30$\degr$ and 60$\degr$ for \mfif\ and between
about 30$\degr$ and 45$\degr$ for \meig.  These imply that the maximum line-of-sight optical depth at 225$\,$GHz
was typically much less than 0.3.  

Interleaved between groups of raster scans of the program galaxies were observations of additional sources to 
focus the {\it AzTEC\/} camera, to check the pointing, and to calibrate the data.  Focusing was achieved through repeated 
jiggle-maps of a few chosen point-like sources to minimize the beam's angular size and maximize its peak.  Focusing was 
done each night usually on the planet Uranus, but also on the late-type stars CRL618 and IRC$+$10216.  Pointing was 
checked and corrected by small jiggle-maps of QSOs 1308$+$326 and 1334$-$127.  The pointing maps were performed before 
and after many raster scans of the program galaxies, or about every 1 to 1.5 hours.  The $rms$ pointing
uncertainty was 2$''$ or better.  Beam
maps of Uranus were made to determine the flux conversion factor which converts the detector output voltages
to $mJy\cdot beam^{-1}$ \citep[see][for details]{Wilson08}.

Spectral lines do not contribute appreciably to the {\it AzTEC\/} 1.1-mm bandpass.  The strongest line in this bandpass
is CO~$\Jtwo$.  This is at the edge of the bandpass for the 1.1-mm filter \citep[see Figure~4 of][]{Wilson08}
and line peak would be attenuated by a factor of $\sim 100$ due to the low response of the filter at this frequency.
In addition, the {\it AzTEC\/} bandwidth at 1.1$\,$mm is about 70$\,$GHz, resulting in a spectral dilution of the line of
a factor of $\sim 10^3$.  The two effects together dilute the line strength by a factor of $\sim 10^5$.  
Publicly available CO~$\Jone$ maps \citep[e.g.,][]{Helfer03, Crosth02} along with adopting a reasonable
ratio for the $\Jtwo$ to $\Jone$ lines (i.e. 0.7)  and 
applying the peak attenuation, the spectral dilution, and the 
observed 1.1-mm fluxes of \mfif\ and \meig\ (see Section~\ref{surf_bright}) yield a relative contribution of  
5-6$\times 10^{-4}$ by the CO~$\Jtwo$ line to 
the total observed 1.1-mm flux.  Including the effects of other spectral lines, even those with peak attenuations
closer to unity, are unlikely to add a total flux contribution of more than a few percent to the {\it AzTEC\/} 1.1-mm band.

\section{Data Reduction}

The emission observed by the telescope in the 1.1$\,$mm continuum is dominated by that of the atmosphere.  
In fact, such atmospheric emission is roughly a factor of 1-$3\times 10^5$ stronger than that from astronomical 
sources.  Hence, considerable data processing is necessary for extracting the faint astronomical signal from 
the time-series data that will be converted into an astronomical image.  This processing assumes that the spatial 
extent of the atmospheric emission is greater than that of the astronomical sources; this makes it possible, though 
still difficult, to reconstruct extended structures in the image.   Recovering the extended astronomical emission
requires an iterative procedure whereby the image from the first iteration is subtracted from the time-series
data and these image-subtracted time-series data are used to construct the next iteration image, which is again
subtracted from the time-series data and so on, until a suitable convergence is achieved.  The algorithm
described here is very similar 
to that for the M$\,$33 observations by 
\citet{Komugi11}, but with more emphasis on recovering the large-scale structure.

The data reduction starts with the full pipeline routine that removes spikes and calibrates the data, and
applies principal component analysis (PCA) to filter out extended emission that largely represents the earth's
atmosphere.  
Included with the software
pipeline output were noise maps that were created by the jack-knifing technique as described in \citet{Komugi11}.  
Iterations or loops of the algorithm, Flux Recovery Using Iterative Technique or FRUIT, are then applied to the
preliminary map in order to recover the missing extended emission.  The FRUITloops subtract a preliminary
map from the time-series data and a new map is created from those data.  
The different iteration maps are examined to find the latest iteration that is free
of artefacts, such as large patches of negative emission.  This is usually the final iteration of FRUITloops.
If no iteration map is acceptable, then the parameter values used in the full pipeline and FRUITloops are
changed and those routines are run again. 

Estimating reasonable parameter values for a reliable reconstruction of the image required creating simulated
images.  The map from the final iteration of FRUITloops was subtracted from the time-stream data and an input 
test map was then added into that time-stream.   The test map for \mfif\ was the 24$\um$ map of \mfif\ from 
\citet{Dale09} and for \meig\ it was the 5$\,$GHz map of \meig\ from \citet{Nein93}.   These test maps were placed on the 
same pixel grid as the corresponding {\it AzTEC\/} maps and were convolved to the 18$''$ resolution of those maps. 
After being added into the time-stream data (now more or less devoid of the {\it AzTEC\/} detected astronomical flux),
the data are processed through the full pipeline and then FRUITloops.  For each iteration of FRUITloops the 
difference between the output map and the input map is used to determine a reduced chi-square, $\rchi$.  If the 
new simulation output map is acceptable, then the data processing is re-run on the real data with the new
parameter values that were used in the latest simulation.  The new real-data map is removed from the time-stream 
data and the simulations are run again.    If, however, there are image artefacts in the simulated maps or the 
$\rchi$ is too large, then the parameter values used 
in the full pipeline and in FRUITloops are changed and the simulation is re-run.  This 
process continues, going from simulations to real-data processing and back again, until the real and simulated
output maps are free of artefacts and the $\rchi$ from the simulations is around 1-2.  

The final {\it AzTEC\/} maps were converted to units of \mjysr and smoothed to a final resolution of 20$''$
so that the final maps were less choppy.   The
conversion factor to units of $mJy\cdot beam^{-1}$ is 8.63 for the original 18$''$ beam.  

With the help of the simulations, the optimum, or near-optimum, parameter values were determined in
the processing of the \mfif\ and \meig\ data sets.   The simulations showed a minimum $\rchi = 2.0$ and $\rchi=0.45$ 
for \mfif\ and \meig, 
respectively.  The final map of noise levels, the sigma map, of \mfif\ was scaled upward by $\sqrt{2}$ to account for the 
$\rchi$ that was greater than unity.  
The large-scale emission was faithfully recovered with some minor problems for both galaxies.  
By differencing the final output map with the initial input map of the simulation, and scaling from simulation output map
to real output map, the large-scale surface brightness of the real map was checked.   

The simulations tell us that the {\it AzTEC\/} \mfif\ map
underestimates the true surface brightness at 
an average of 0.3\mjysr, or the $rms$ noise level for much of the map.  
This has a trivial effect locally, but has a non-trivial effect on the large scale.   Accordingly, the offset determined here was not added to 
the final \mfif\ map, but the simulation input and output maps were still used to check any 
results derived from the \mfif\ {\it AzTEC\/} map.   


The case of \meig, however, was quite different.   
Averaged over the entire central
12.2$\,kpc$ radius, the true surface brightness is 0.05\mjysr or 0.25~$\sigma$ {\it lower\/} than that of the map.  This
correction is small enough that using the simulation to correct the results derived for \meig\ was unnecessary.

The input and output maps of simulations for \meig\  were also compared visually and no strong artefacts were found. 

\begin{figure*}
\vspace{-25mm}
\includegraphics[height=225mm, keepaspectratio=true]{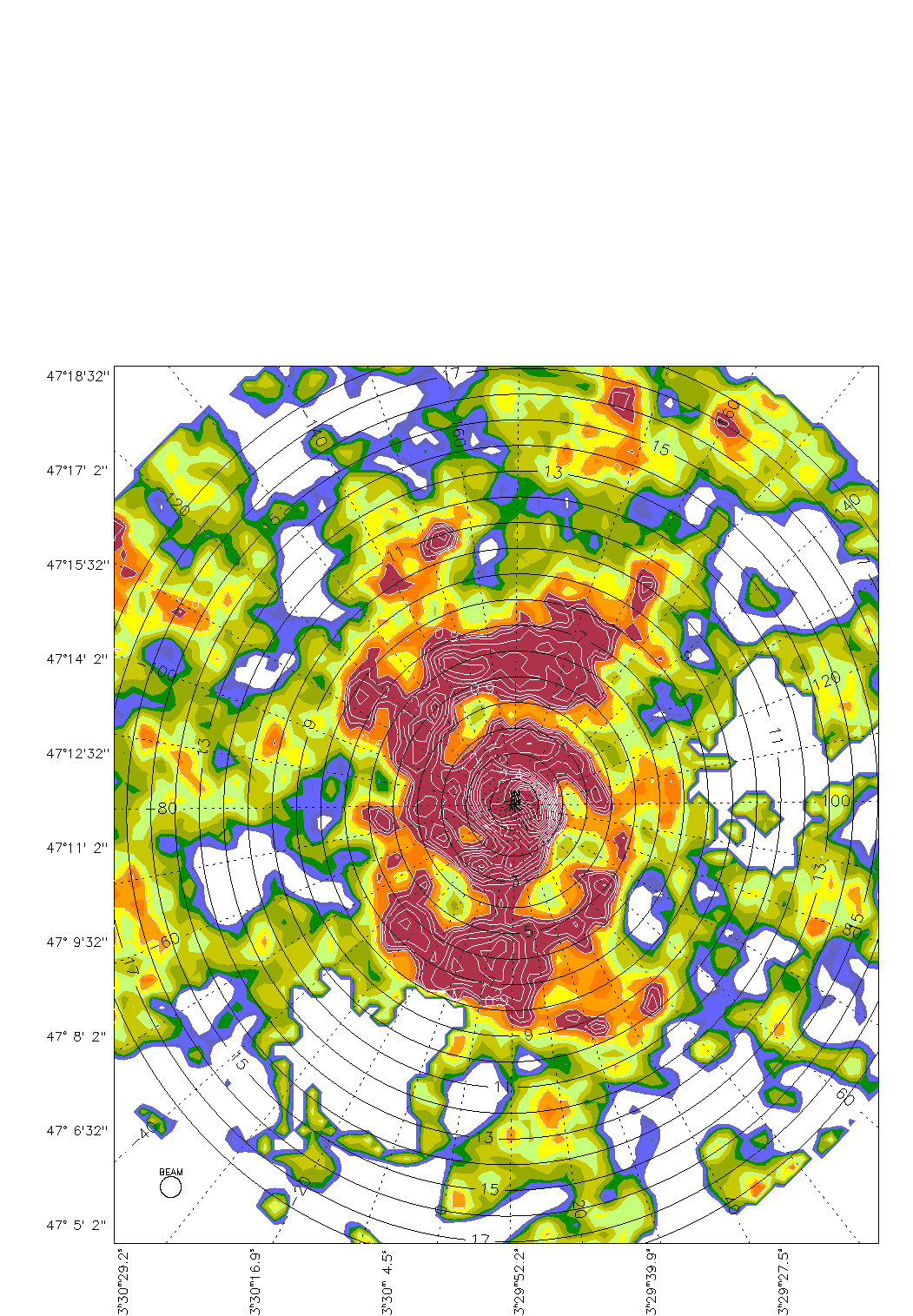}
\vspace{2mm}
 \caption{\mfif\ continuum map at wavelength 1.1$\,$mm.  The coordinates are epoch 2000.0.  The solid white line
contour levels are 0.6, 0.8, 0.9, 1.0, 1.2, 1.4, ..., 3.8\mjysr.  The concentric ellipses represent distances in 1$\,$kpc
increments from the adopted centre position in the plane of \mfif\ for its adopted distance.  The dashed lines represent 
position angles relative to the major axis.  See Table~\ref{tab1} for details. The effective beam is
shown in the lower left corner.}
 \label{fig1}
\end{figure*}

\begin{figure*}
\vspace{-5mm}
\includegraphics[height=225mm, keepaspectratio=true]{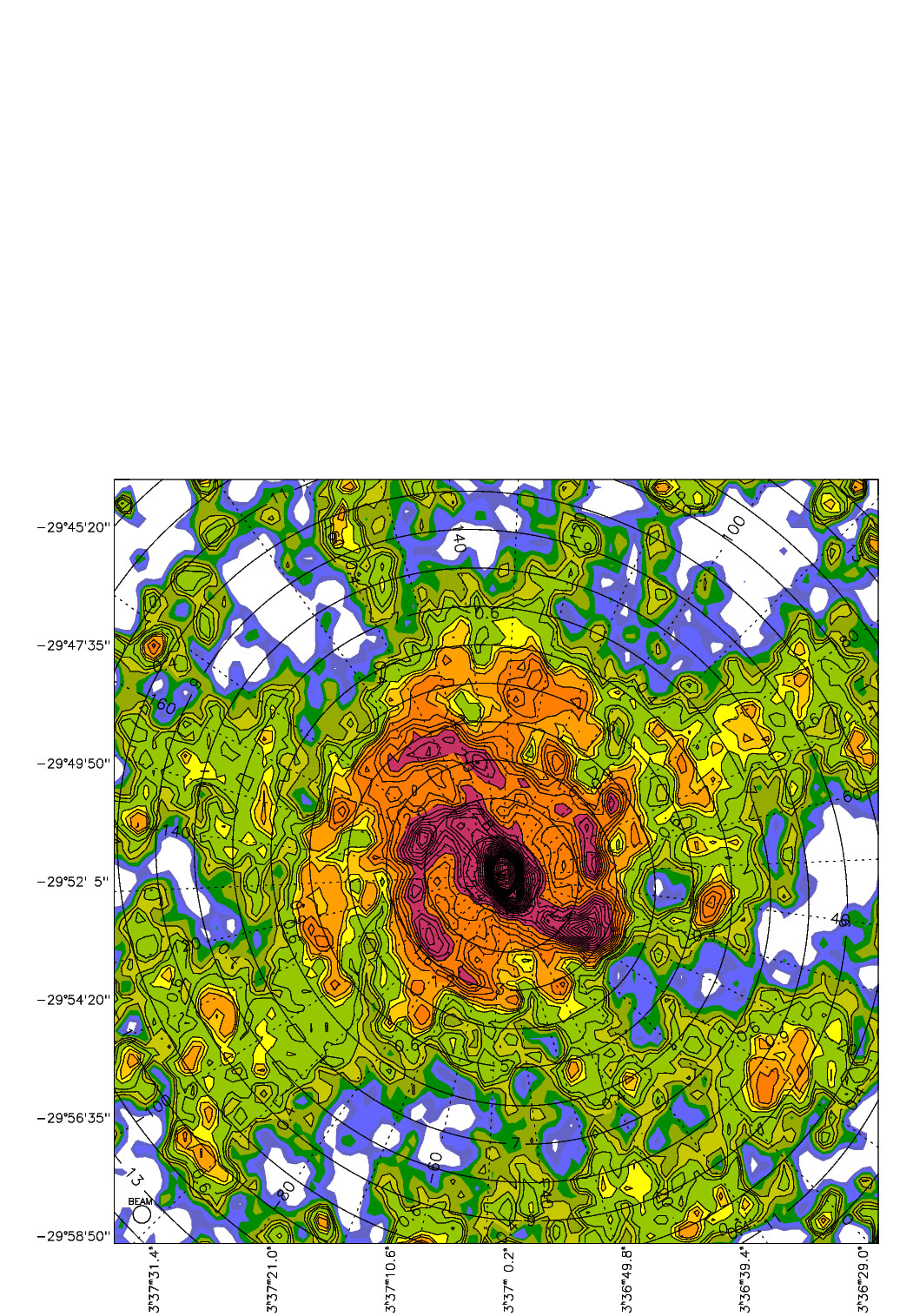}
\vspace{-1mm}
 \caption{\meig\ continuum map at wavelength 1.1$\,$mm.  The coordinates are epoch 2000.0.  The solid line
contour levels are 0.4, 0.5, 0.6, 0.8, 0.9, 1.0, 1.2, 1.4, ..., 7.2, 8.0, 9.0, 10.0, 11.0, 12.0\mjysr. The concentric 
ellipses represent distances in 1$\,$kpc increments from the adopted centre position in the plane of \meig\ for 
its adopted distance.  The dashed lines represent position angles relative to the major axis.  See Table~\ref{tab1} 
for details.  The effective beam is shown in the lower left corner.}
 \label{fig2}
\end{figure*}

\begin{figure*}
\vspace{-25mm}
\includegraphics[height=225mm, keepaspectratio=true]{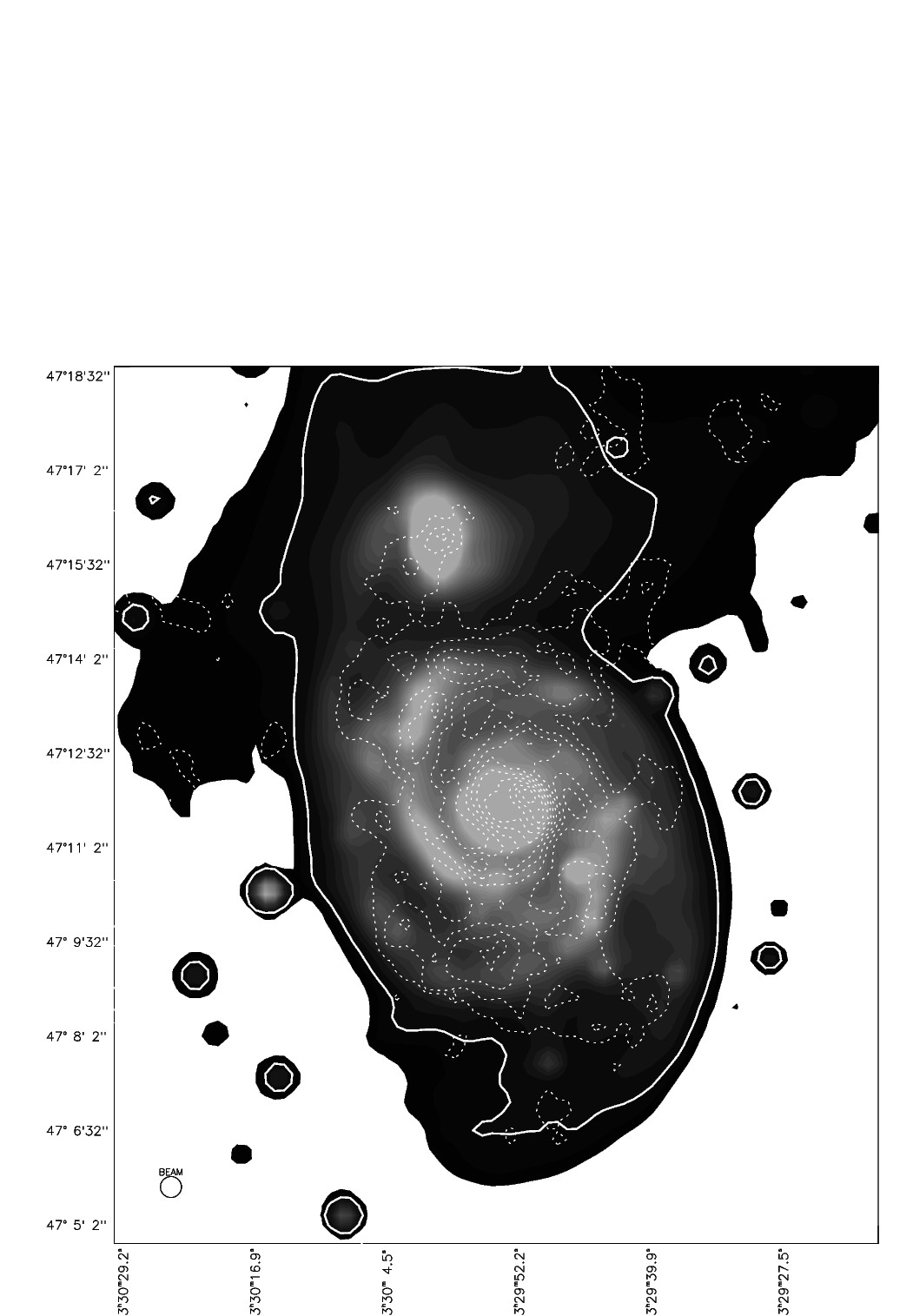}
\vspace{2mm}
 \caption{\mfif\ continuum map at wavelength 1.1 mm (white, dashed contours) is superposed on an image at 
visible wavelengths \citep[Sloan g-band, see][]{Baillard11} that has been smoothed to the resolution
of the {\it AzTEC\/} observations. The solid white contour represents the 
25 $mag\cdot arcsec^{−2}$ isophote. The effective beam is shown in the lower left corner.
}
 \label{fig1a}
\end{figure*}

\begin{figure*}
\vspace{-5mm}
\includegraphics[height=225mm, keepaspectratio=true]{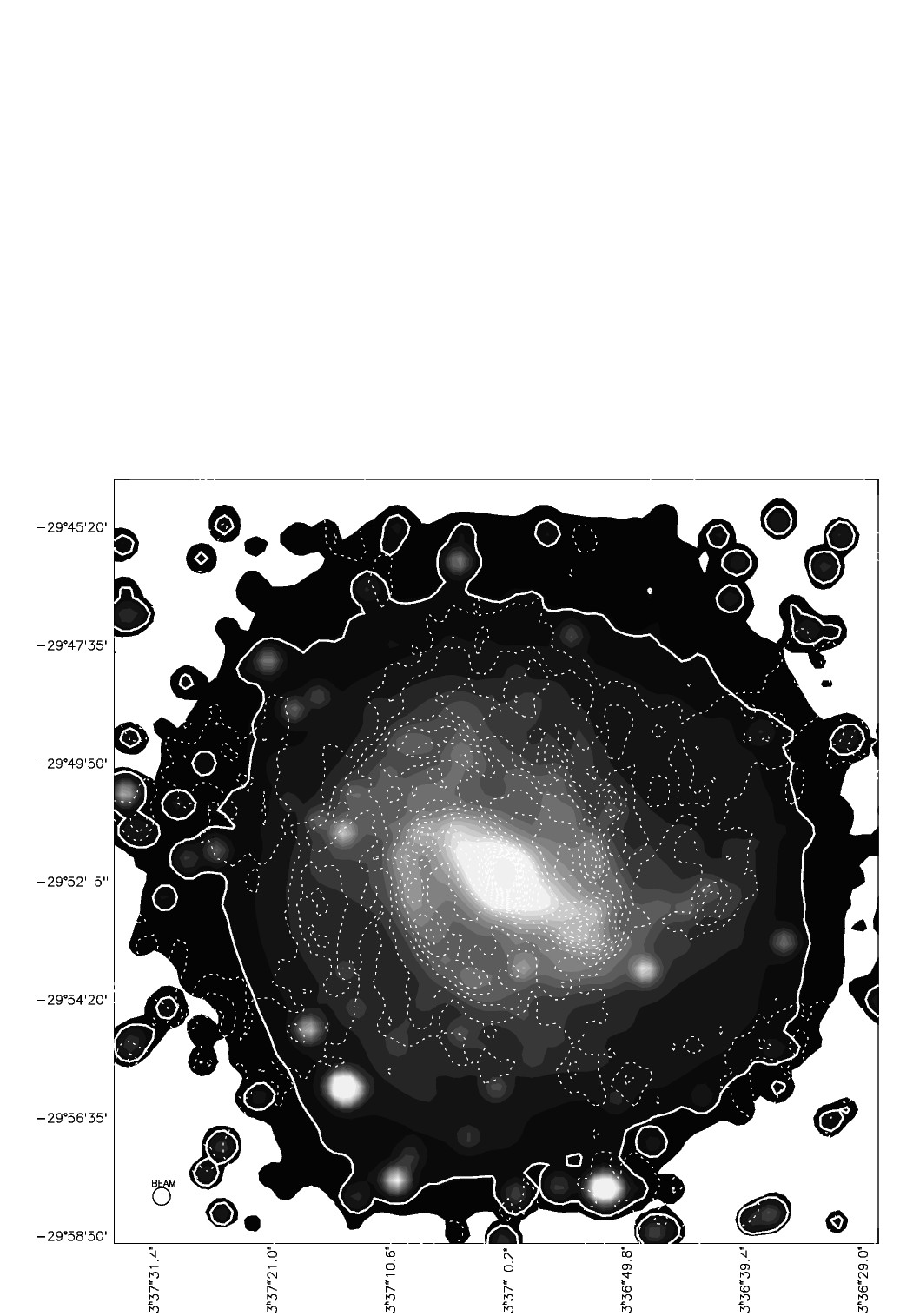}
\vspace{-1mm}
 \caption{\meig\ continuum map at wavelength 1.1 mm (white, dashed contours) is superposed on an image at 
visible wavelengths \citep[R-band, see][]{Meurer06} that has been smoothed to the resolution
of the {\it AzTEC\/} observations. The solid white contour represents the 
25 $mag\cdot arcsec^{−2}$ isophote. The effective beam is shown in the lower left corner.
}
 \label{fig2a}
\end{figure*}

The final maps are shown in Figures~\ref{fig1} and \ref{fig2}.  Comparisons with images
at visible wavelengths are given in Figures~\ref{fig1a} and \ref{fig2a}.
 
\clearpage

\clearpage

\clearpage

\clearpage

\clearpage

\section{Results}

\subsection{Surface Brightness Distribution and Flux}\label{surf_bright}

As is clearly seen in both Figures~\ref{fig1} and \ref{fig2}, both galaxies have extended low-level 
emission in addition to two prominent spiral arms.  In the \meig\ image, the structure visible within
a 1$\,$kpc radius of the centre represents \meig's bar.  The total derived fluxes are $5.6\pm 0.7\, Jy$ for \mfif\ and $9.9\pm 1.4\,Jy$ for \meig.   
(See Appendix~\ref{appflux} for more details.)  For the adopted distances, the
luminosity at 1.1$\,$mm is $L_\nu(1.1\,mm) = (4.7\pm 0.6)\times 10^{29}\, erg\cdot s^{-1}\cdot Hz^{-1}$
or $\nu L_\nu(1.1\,mm) = (3.1\pm 0.4)\times 10^7\, L_\odot$ for \mfif.  For \meig, these are  
 $L_\nu(1.1\,mm) = (2.4\pm 0.3)\times 10^{29}\, erg\cdot s^{-1}\cdot Hz^{-1}$ or 
$\nu L_\nu(1.1\,mm) = (1.7\pm 0.2)\times 10^7\, L_\odot$. 

\begin{figure}
\vspace{0mm}
\includegraphics[height=100mm,angle=90]{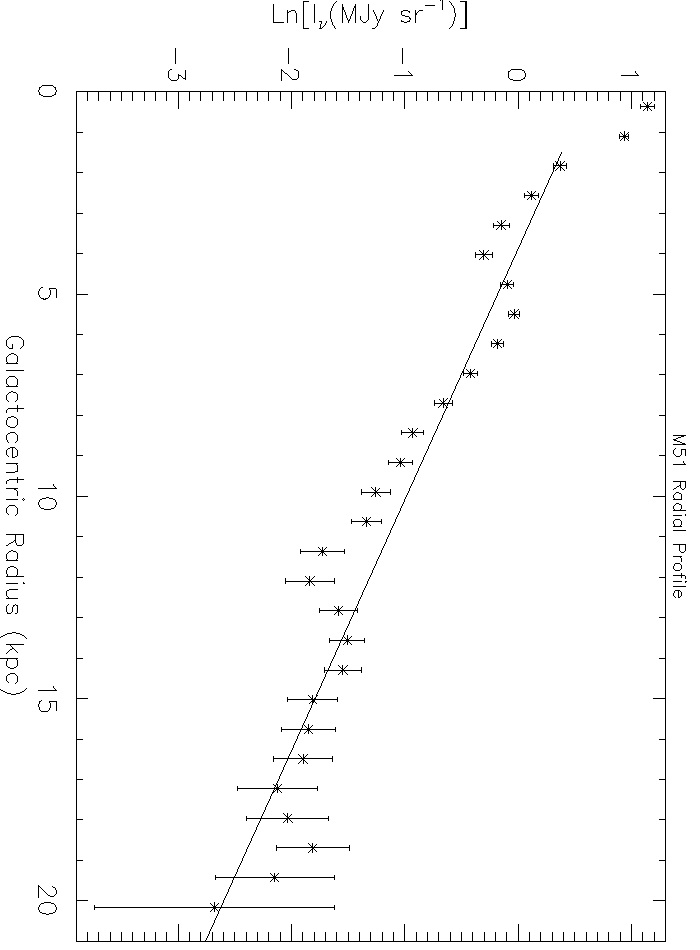}
\vspace{-4mm}
 \caption{The \mfif\ radial profile in the continuum at wavelength 1.1$\,$mm.   The natural logarithm of the
1.1$\,$mm surface brightness of the azimuthally averaged image is plotted against the galactocentric radius in 
kiloparsecs.  The averages are determined within concentric annuli where
each annulus is 2 pixels wide (18$''$ or 0.73$\,$kpc).  The solid line is a linear fit to the logarithms of the
surface brightnesses for radii outside of the central region (see text for details). }   
 \label{fig3}
\end{figure}

\begin{figure}
\vspace{-5mm}
\includegraphics[height=100mm,angle=90]{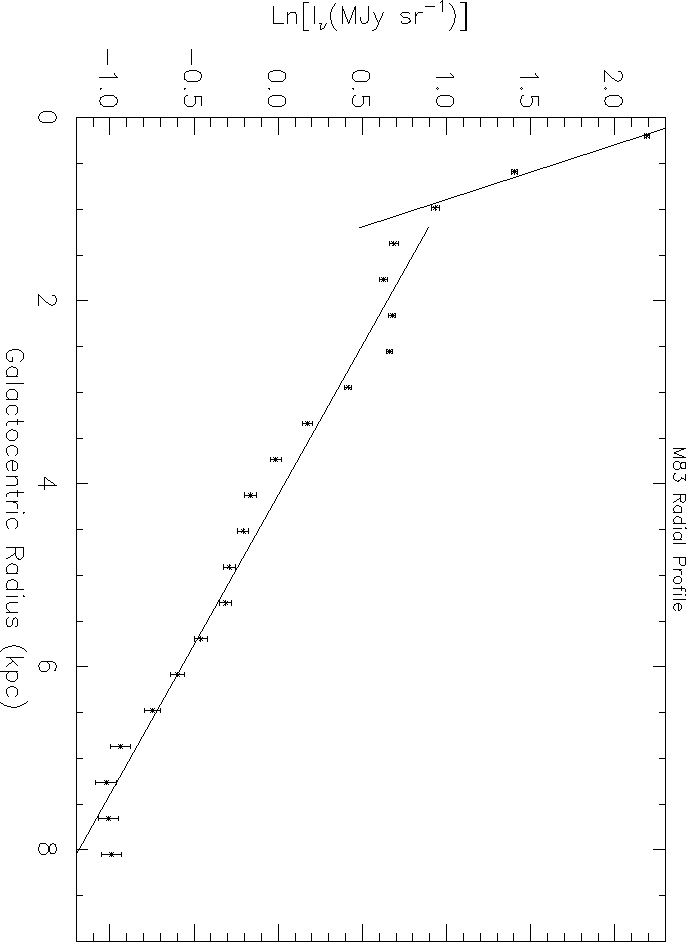}
 \caption{The \meig\ radial profile in the continuum at wavelength 1.1$\,$mm.   The natural logarithm of the
1.1$\,$mm surface brightness of the azimuthally averaged image is plotted against the galactocentric radius in 
kiloparsecs.  The averages are determined within concentric annuli where
each annulus is 2 pixels wide (18$''$ or 0.39$\,$kpc).  The solid lines are linear fits to the logarithms of the
surface brightnesses for radii inside and outside of the central region (see text for details).}
 \label{fig4}
\end{figure}

On the large-scale, each galaxy possesses an exponential disk, as is seen in Figures~\ref{fig3} and \ref{fig4}, where
the non-trivial deviations from the exponential fit represent the overlying spiral structure. 
The exponential scale-lengths are given in Table~\ref{tab2}
are given for different tracers of galactic structure, including that from the current work. 
For \meig, the CO~$\Jone$, near-IR, and 1.1-mm continuum have scale-lengths that are comparable to within 
1- or 2-$\sigma$.   For \mfif, however, the 1.1$\,$mm exponential disk is more than 5-$\sigma$ larger than both the stellar 
disk (as represented by the near-IR) and the disk of CO~$\Jone$ emission.  Even excluding the correction factor derived 
from the simulations would reduce this difference by very little. \citet{Meij05} estimated a scale-length for the dust from 
850$\um$ observations that is
consistent with 1.1-mm observations. 

\begin{figure*}
\vspace{-20mm}
\includegraphics[height=245mm, keepaspectratio=true]{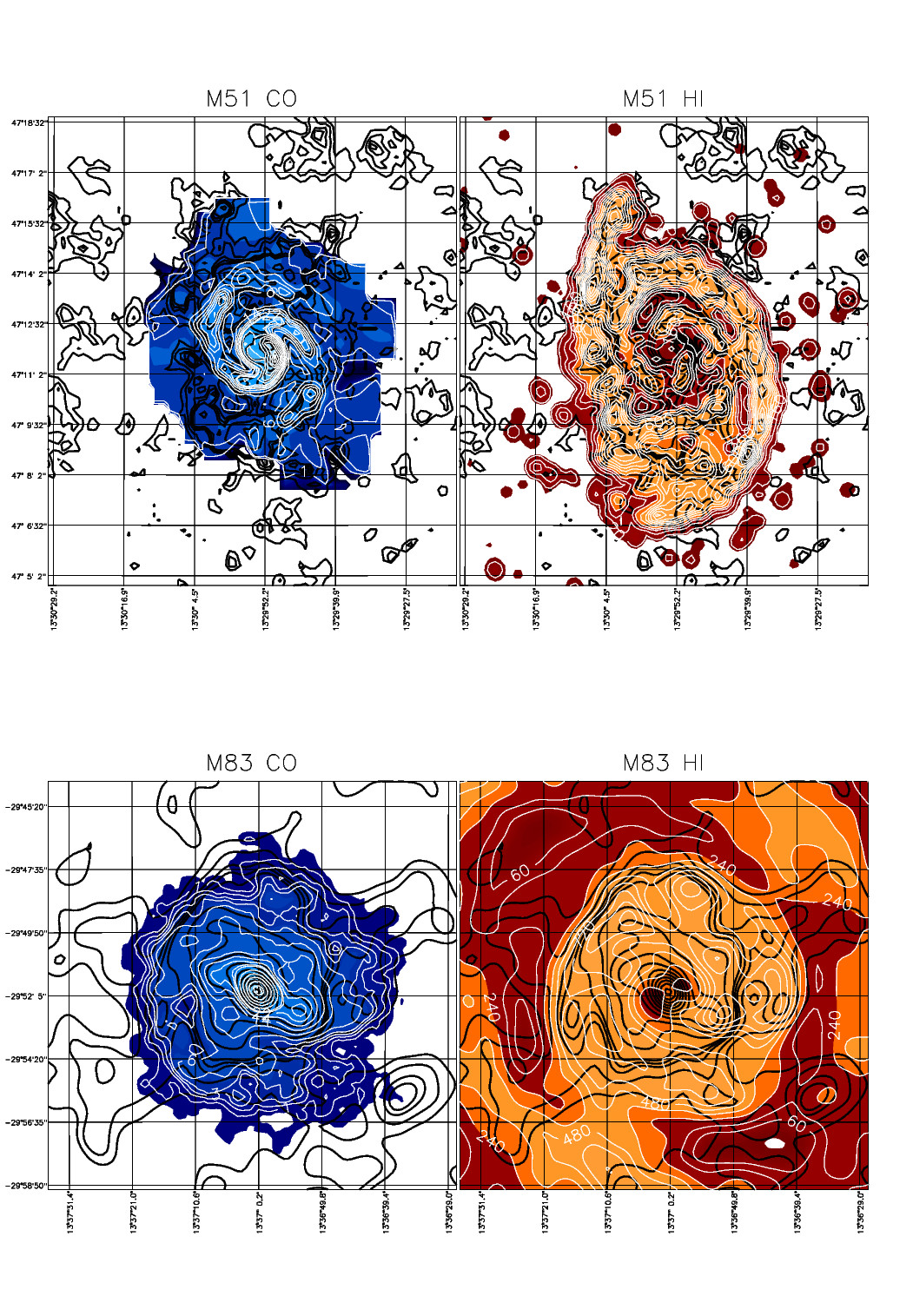}
\vspace{-16mm}
 \caption{CO $\Jone$ and HI maps are superposed on the corresponding 1.1$\,$mm continuum maps of \mfif\ and \meig.
The 1.1$\,$mm continuum map in each panel is represented by the dark contours with levels of 0.4, 0.6, 0.8, 0.9,
1.0, 1.4, 1.8, 2.2, ..., 3.8\mjysr for \mfif\ with the same sequence of levels for \meig, but with a maximum level of 
5.0\mjysr.  The 1.1$\,$mm spatial resolution is 20$''$ in the \mfif\ panels and 55$''$ in the \meig\ panels. In the 
\mfif\ CO (upper left) panel, the white contours and blue shading show the CO~$\Jone$ emission convolved to a 
20$''$ resolution with contour levels of 1, 2, 3, 4, 6, 10, 14, 18, 22, ..., 76$\Kkms$.  In the \mfif\
HI panel, the white contours and red/orange shading show the HI emission convolved to a 20$''$ resolution with
contour levels of 20, 80, 160, 240, 360, 480, ..., 1560$\Kkms$.  The \meig\ CO panel displays the CO~$\Jone$ image
at a 55$''$ resolution with the same sequence of contour levels as for the \mfif\ image, but with 68$\Kkms$ as the
maximum level. The \meig\ HI panel depicts the HI emission convolved to 55$''$ resolution with  white contours 
and red/orange shading; the contour levels are 60, 120, 240, 360, 480, ..., 1320$\Kkms$.   The lowest contour level
is roughly equivalent to 1$\,\sigma$ for all maps in this figure. 
}
 \label{fig2aa}
\end{figure*}

The 1.1$\,$mm maps of \mfif\ and \meig\ are very similar to their corresponding CO~$\Jone$ and HI~21-cm maps, as we
can see in Figure~\ref{fig2aa}.  For example, the CO maps of both galaxies are similar to that of the
1.1$\,$mm continuum out to a radius of about 6$\,$kpc.
Beyond that radius, the millimetre-continuum emission extends further than the CO~$\Jone$ emission, especially 
in the northern arm of \mfif.
That more extended emission is partly due to dust associated
with atomic gas.  In \meig, there is a bridge of millimetre-continuum emission extending out to 9$'$ or about 12$\,$kpc 
from the nucleus to both the southeast and to the northwest.  This bridge is also seen in HI, but is shifted counterclockwise
in position-angle by about 15$\degr$ with respect to the millimetre-continuum bridge, where this shift is likely
an artefact of the missing large-scale HI emission in the interferometer map.  



\begin{table}
 \caption{Exponential Disk Scale-Lengths for Different Tracers$\, ^a$}
 \label{tab2}
 \begin{tabular}{@{}lll}
  \hline
  Tracer & \mfif & \meig \\
  \hline
  Near-IR & $1.8\pm 0.2\, ^{b,c}$ & $2.4\pm 0.4\, ^{d,e}$ \\
  CO $\Jone$ & $2.8\pm 0.1\, ^c$ & $2.9\pm 0.3\, ^e$ \\
  1.1$\,$mm  & $7.3\pm 0.8\, ^f$ & $3.3\pm 0.3\, ^f$ \\
  \hline
 \end{tabular}

 \medskip
$^a$ All scale-lengths in kiloparsecs for adopted distances given in Table~\ref{tab1}.

$^b$ 3.6$\um$.

$^c$ \citet{Regan06}. 

$^d$ 2.2$\um$.

$^e$ \citet{Lundgren04}.

$^f$ Excluding central 1.5$\,$kpc radius for \mfif\ and 1.2$\,$kpc radius for \meig.

\end{table}

\clearpage 
\clearpage

\subsection{Surface Density Distribution and Mass}\label{surfdens}

The 1.1$\,$mm continuum surface-brightness maps of Figures~\ref{fig1} and \ref{fig2} can be converted
to maps of surface density, or column density, if the dust temperature is known at each position.   Hence, the 
{\it AzTEC\/} 1.1$\,$mm maps were ratioed with the {\it Spitzer/MIPs\/} 160$\um$, effectively 155.9$\um$, 
maps to yield these temperatures.   The derived column densities were calibrated against the HI gas column densities
at those positions where molecular appears to not dominate. This approach was modified by removing upper outliers
resulting in the intermediate-$\knu$ case adopted for the current paper. (See Appendix~\ref{appsurf} for details.)  

Figures~\ref{fig5} to \ref{fig6a} display the derived dust temperature maps, the column density maps, and their
radial profiles for both \mfif\ and \meig.  
Both the radial $\Td$ profile of \mfif\ in Figure~\ref{fig5a} and that of \meig\ in
Figure~\ref{fig6a} have relatively constant temperatures of $\sim 20$-$25\,$K out to a radius of 3-4$\,$kpc, a linear decline out
to 15$\,$kpc for \mfif\ and 9$\,$kpc for \meig, and then a more or less constant temperature of $\sim 12\, K$ in the outer disk.

\citet{Foyle12} had data at five wavelengths and could produce maps of both dust temperature and $\beta$.  
The apparently lower dust temperature in the spiral arms compared to that of the interarm dust is merely an artefact
of not accounting for the spatial variation of $\beta$.   
In general, the dust 
temperatures determined from the 1.1$\,$mm data are $\sim 5\,$K lower than those determined from the shorter
wavelengths 
of the {\it HERSCHEL\/} data.  This suggests that the millimetre
continuum is sampling an additional component of the dust.   The dust mass derived in the current work,
using their dust-to-gas mass ratio and dust mass absorption coefficient (equivalent to our low-$\knu$ case)
is a factor of 2.5 higher than their $4\times 10^7\,M_\odot$ over the equivalent area.
This is roughly consistent with the lower $\Td$ values that we derive.  Also, based on the {\it AzTEC/SPITZER\/} data alone, the area covered 
by the \citet{Foyle12} map is sampling half the dust and gas mass of \meig.

A similar comparison between the {\it HERSCHEL\/} \mfif\ observations \citep{Cooper12} and those of {\it AzTEC\/}
yield nearly identical masses ($<1\%$ difference) for the \citet{Cooper12} NGC$\,$5194 field, after adjusting to their
adopted distance, to the low-$\knu$ case, and to their gas-to-dust mass ratio.  
Repeating this for their NGC$\,$5195 field, however, yields a 
disagreement of a factor of 4.  This can be at least partly attributed to the differences between their observed
dust temperatures and ours.   They find $\Td\sim 35\,K$, whereas we find $\Td\sim 22\,K$.  These temperatures 
roughly correspond to the {\it AzTEC/SPITZER\/} data yielding a dust mass that is a factor of 4 higher.  As was
the case for \meig, the 
longer wavelength {\it AzTEC\/} observations appear to be sampling an additional component.   Interestingly, this does not seem to
be the case for the NGC$\,$5194 field, where agreement is very tight.   Also, based on the {\it AzTEC/SPITZER\/} 
data alone, the area covered by the \citet{Cooper12} map is sampling one-third the dust and gas mass of \mfif.  If
the constant-offset correction determined from the simulations is not applied, then this one-third is a less
extreme one-half.

Despite the differences found for each of the \meig\ and for the NGC$\,$5195 fields, it must be emphasized that 
the NGC$\,$5194 field 
has the same mass for  both data sets (after applying the appropriate corrections mentioned previously); this strongly 
suggests that the millimetre-wavelength data are not always necessary for a reasonable dust mass estimate and also that 
having only two continuum wavelengths will give reasonable mass estimates (i.e. dust-inferred gas mass), especially when 
calibrated against gas column densities. 

It should also be mentioned that the companion galaxy, NGC$\,$5195, contributes very little to the total
masses or fluxes: $\sim$5\%.  Its presence can be largely ignored. 

Figures~\ref{fig5a} and \ref{fig6a} demonstrate detailed agreement to within factors of about 2 
between the spectral line derived and continuum derived column densities out to a galactocentric radius 
of about 6$\,$kpc for both galaxies.  
Beyond a radius of about 
8$\,$kpc, the spectral line derived column densities are a factor of 2 or more below those derived from the 
continuum.  This is possible evidence that the X-factor is strongly rising in the outer disks of both galaxies,
although other explanations are not entirely ruled out.  This will be addressed in more detail in Sections~\ref{Xfac} and \ref{xfsv}.

The total mass of gas, $M_d(H)$, in \mfif\ out to a radius of 13.6$\,$kpc is $9.2\times 10^9\, M_\odot$, 
including the correction given by the simulations (or $6.6\times 10^9\, M_\odot$ without this correction).   
For \meig, $M_{gas}$ is $7.2\times 10^9\, M_\odot$ out to a radius of 12.2$\,$kpc. 
For the adopted parameters (i.e., distance, intermediate-$\knu$, etc.), the observed 1.1$\,$mm flux and the derived mass 
imply	 average dust temperatures of 19.4$\,$K and 13.9$\,$K for \mfif\ and \meig, respectively.  
For both galaxies, 
\begin{equation}
\hfill {M_d(H)\over\nu L_\nu(1.1\,mm)} = (3.6\pm 0.6)\times 10^2 M_\odot/L_\odot , \hfill
\end{equation}  
where the uncertainty represents the range of values for this sample of two.
Using the abovementioned dust temperatures and assuming $\beta=2.0$, the 
 $M_d(H)/\nu L_\nu(500\um)$ is roughly $29\pm 8\, M_\odot/L_\odot$.  This is consistent with \citet{Groves15},
 who found that ratio to be 20-$30 M_\odot/L_\odot$ (as inferred from their Table~8 for massive galaxies).

The continuum derived and spectral line derived masses and a breakdown of gas masses in \mfif\ and \meig\ as well as a comparison between the two are given in Table~\ref{tab3}.  


\clearpage
\begin{figure*}
\includegraphics[height=205mm, angle=0, keepaspectratio=true]{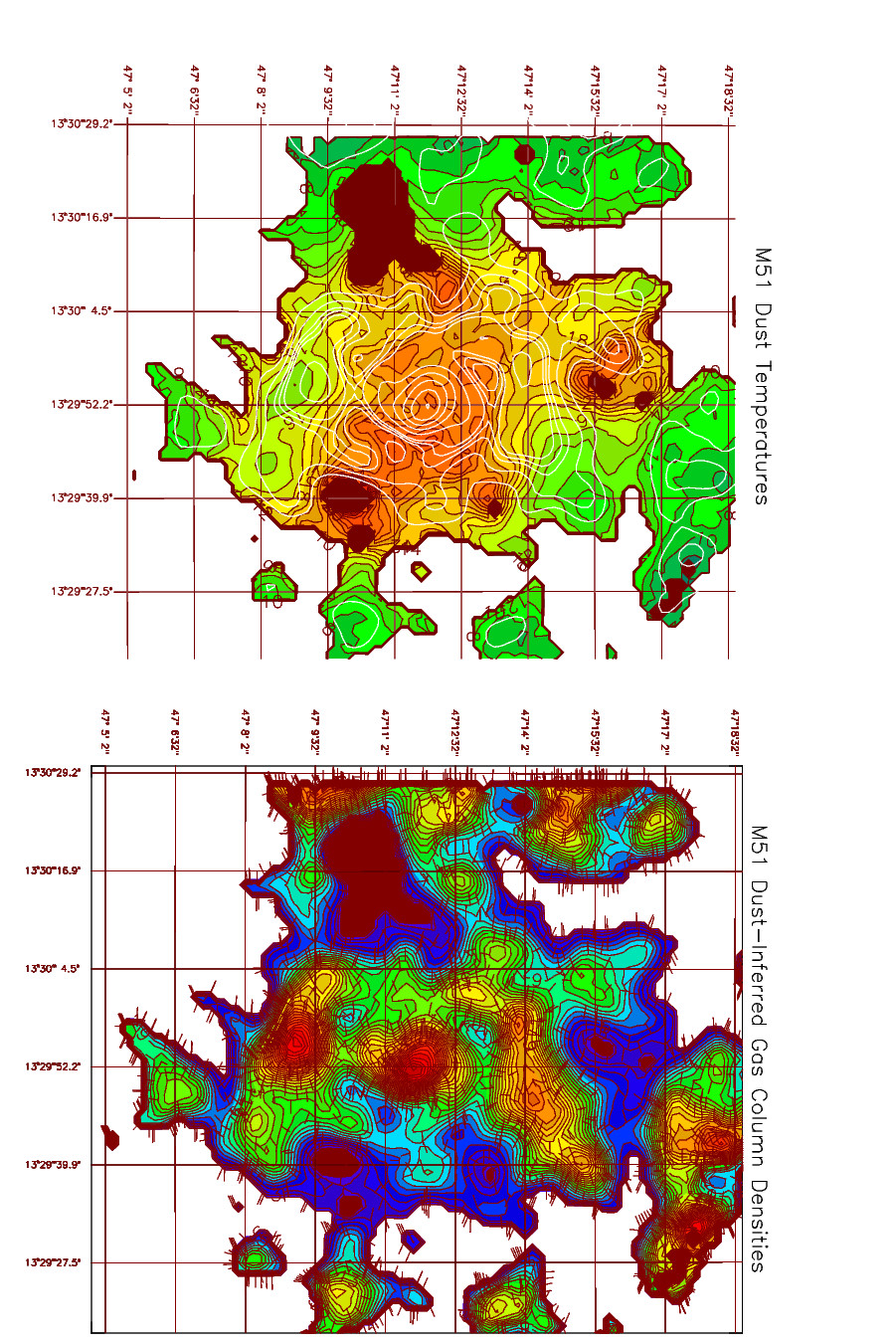}
\vspace{5mm}
 \caption{\mfif\ dust temperature map and H-nuclei column density map are shown in the left and right panels,
respectively. The coordinates are epoch 2000.0.  Left Panel: The dust temperatures are those determined from the ratio
of the {\it Spitzer\/} 156$\um$ to {\it AzTEC\/} 1.1$\,$mm intensity ratio.  The coloured contours give the dust temperatures
in 1$\,$K steps from 8$\,$K.  The dark solid areas represent regions where the signal-to-noise of the intensity ratio was less
than unity.  The white contours give the 1.1$\,$mm intensities in \mjysr for the 1.1$\,$mm continuum
map degraded to the 38$''$ resolution of the 156$\um$ map.  The 1.1$\,$mm contour levels are 0.4, 0.6, 0.8, 0.9, 1.0, 1.4,
1.8, 2.2, 2.6, 3.0\mjysr.  Right Panel: These are gas column densities inferred from the dust-continuum emission (see text).  
The contour levels are 0.5, 1.0, 1.5, 2.0, ,..., 20.0 $\times 10^{20} H$-$nuclei\cdot cm^{-2}$.  
The tick marks point towards lower contour levels.   The dark solid areas correspond to those of the dust temperature map 
in the left panel.}
 \label{fig5}
\end{figure*}

\clearpage
\begin{figure*}
\vspace{0mm}
\includegraphics[height=198mm,width=120mm,angle=0]{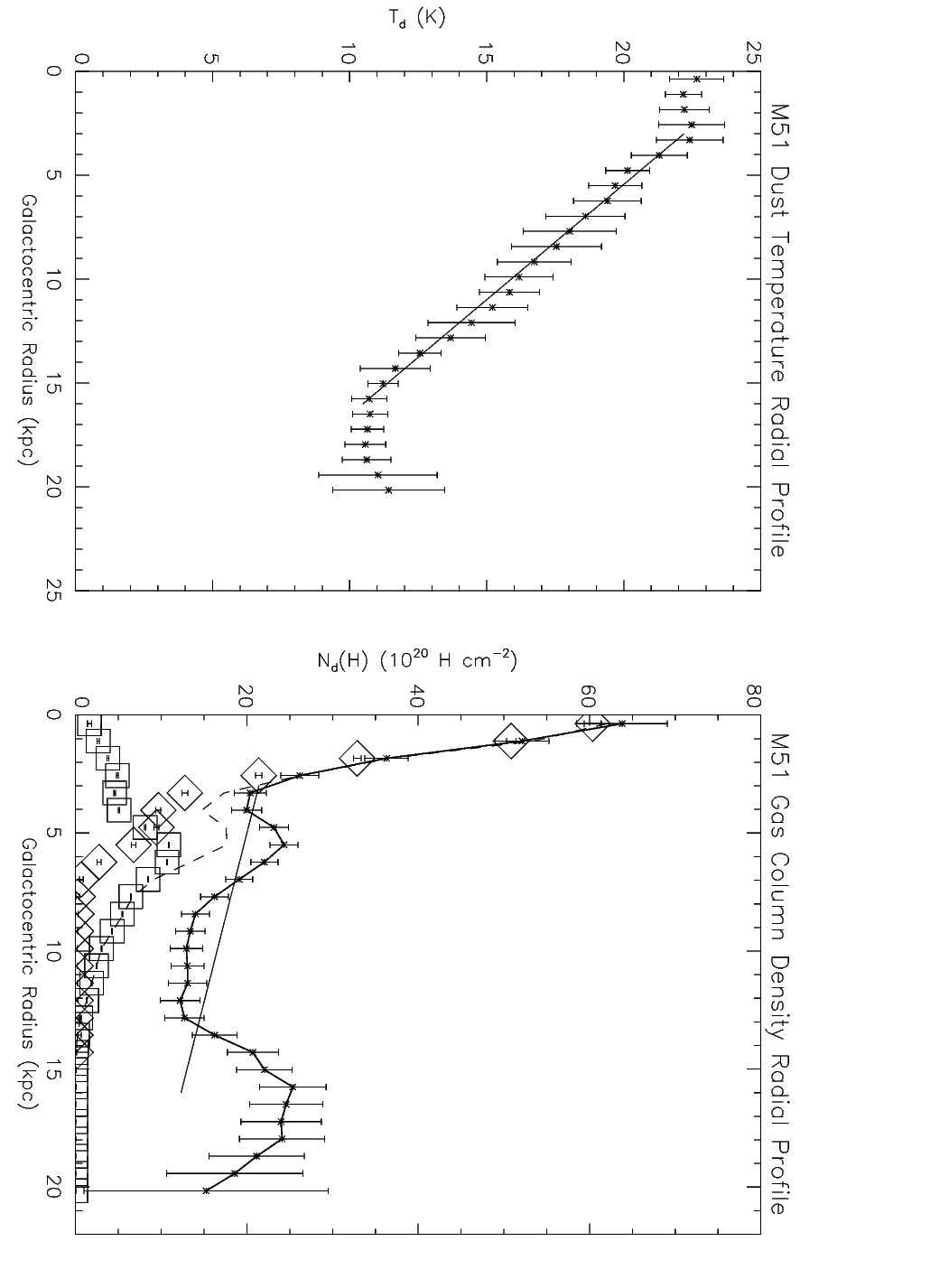}
\vspace{0mm}
 \caption{The \mfif\ radial profile of the 156$\um$/1.1$\,$mm dust temperature and of the H-nuclei column densities, $\Nd$, as 
inferred from the dust continuum emission are shown in the left and right panels, respectively.  The azimuthally averaged
dust temperature and column densities are plotted against the galactocentric radius in kiloparsecs. The averages are determined 
within concentric annuli where each annulus is 2 pixels wide (18$''$ or 0.73$\,$kpc).   The solid line in the left panel is a 
linear fit to radii from 3 to 16$\,$kpc, where the dust temperature has a linear decline of about 0.9$\, K\cdot kpc^{-1}$.   
The radial profile of the H-nuclei column densities is given in the right panel above by the thick solid curve joining the data points.   
The thin solid line in the right panel is a linear fit to the column densities at radii from 3 to 16$\,$kpc, which corresponds 
to where the dust temperature 
has a linear decline.  The diamonds represent the column densities of H-nuclei in the molecular gas as inferred from CO $\Jone$ using a 
constant X-factor (see Section~\ref{Xfac}).  The squares represent the column densities of H-nuclei 
in the atomic gas as inferred from HI 21-cm emission. For both the CO and HI, the error bars are smaller than the symbols.  The dashed line 
gives the total gas column density as inferred from both CO and HI.   Note that the additive correction of 
$5.5\times 10^{20} H$-$nuclei\cdot cm^{-2}$ was not applied to the $\Nd$ points above.
}   
 \label{fig5a}
\end{figure*}

\clearpage
\begin{figure*}
\vspace{-5mm}
\includegraphics[height=205mm, angle=0, keepaspectratio=true]{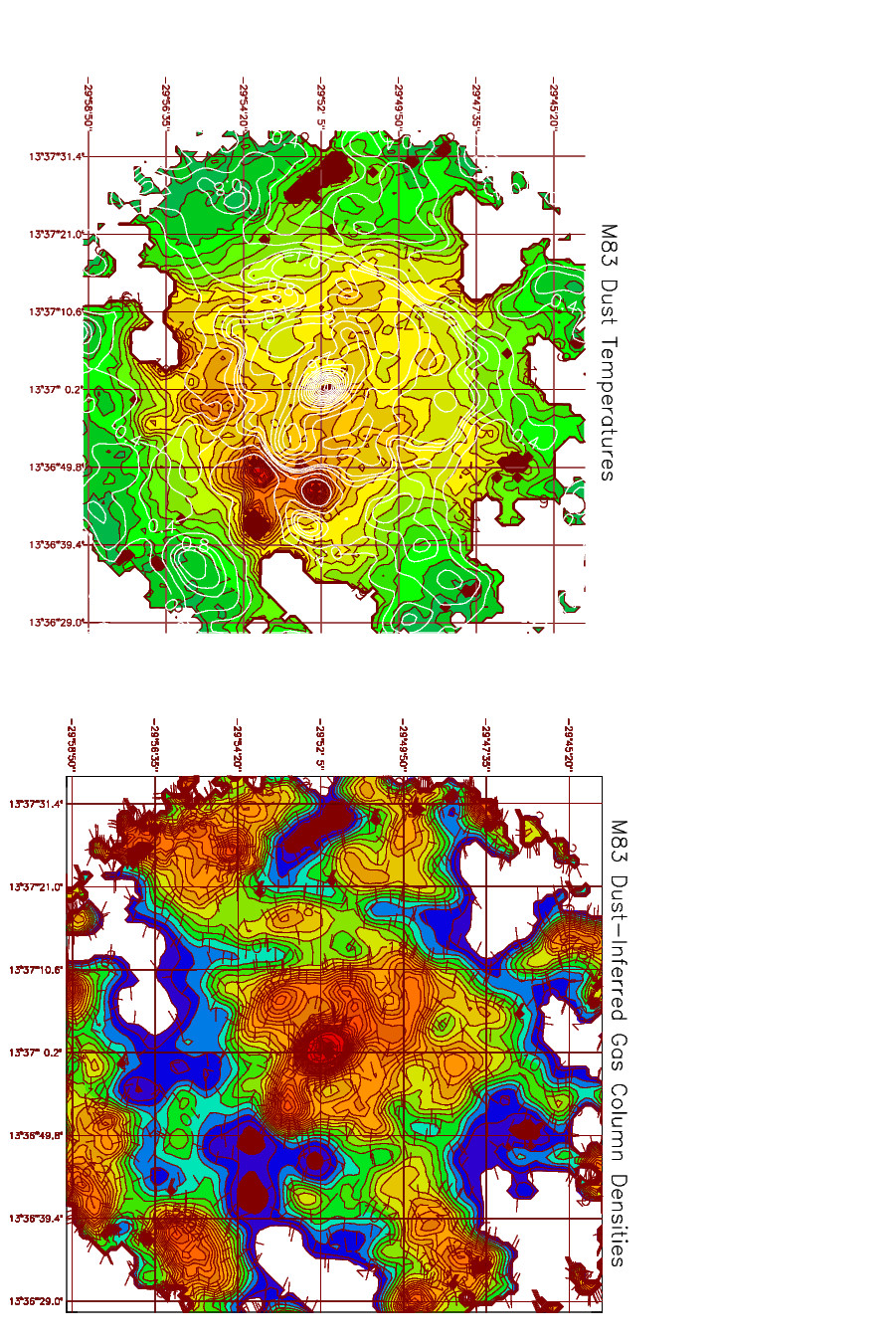}
\vspace{0mm}
 \caption{\meig\ dust temperature map and H-nuclei column density map are shown in the left and right panels,
respectively.  The coordinates are epoch 2000.0.  Left Panel: The dust temperatures are those determined from the ratio
of the {\it Spitzer\/} 156$\um$ to {\it AzTEC\/} 1.1$\,$mm intensity ratio.  The coloured contours give the dust temperatures
in 1$\,$K steps from 9$\,$K.  The dark solid areas represent regions where the signal-to-noise of the intensity ratio was less
than unity.  The white contours give the 1.1$\,$mm intensities in \mjysr for the 1.1$\,$mm continuum
map degraded to the 38$''$ resolution of the 156$\um$ map.  The 1.1$\,$mm contour levels are 0.4, 0.6, 0.8, 0.9, 1.0, 1.4,
1.8, 2.2,..., 7.0\mjysr.  Right Panel: These are gas column densities inferred from the dust-continuum
emission (see text).  The contour levels are 1, 2, 3, 4,..., 30, 35, 40, 45, 50 $\times 10^{20} H$-$nuclei\cdot cm^{-2}$.  
The tick marks point towards lower contour levels.  The dark solid areas correspond to those of the dust temperature map 
in the left panel.}
 \label{fig6}
\end{figure*}

\clearpage
\begin{figure*}
\vspace{-10mm}
\includegraphics[height=198mm,width=120mm,angle=0]{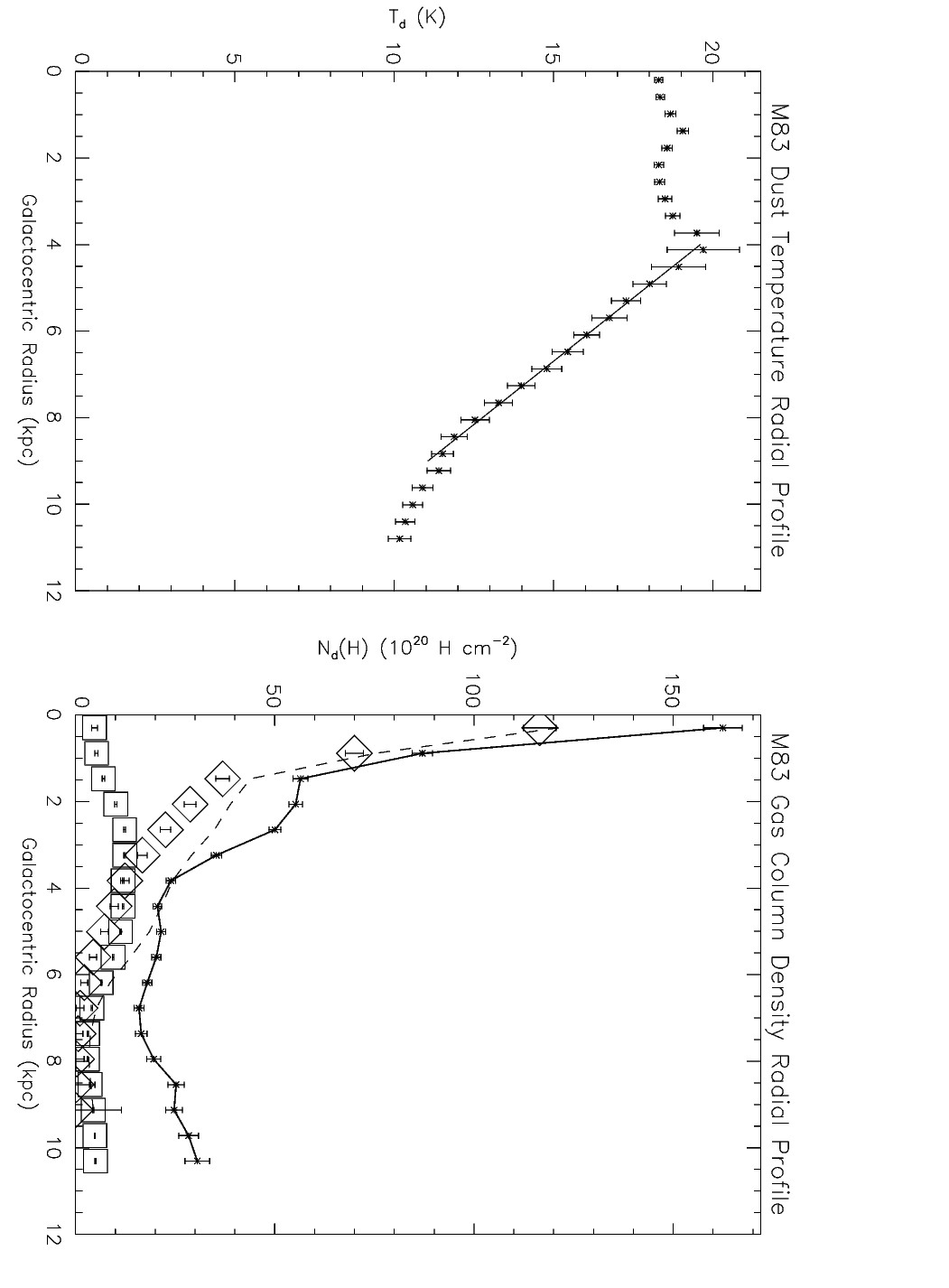}
\vspace{0mm}
 \caption{The \meig\ radial profile of the 156$\um$/1.1$\,$mm dust temperature and of the H-nuclei column densities, $\Nd$, as 
inferred from the dust continuum emission are shown in the left and right panels, respectively.  The azimuthally averaged
dust temperature is plotted against the galactocentric radius in kiloparsecs. The averages are determined
within concentric annuli where each annulus is 2 pixels wide (18$''$ or 0.39$\,$kpc).   The solid line in the left panel 
is a linear fit to radii from 4 to 9$\,$kpc, where the dust temperature has a linear decline of about 1.7$\, K\cdot kpc^{-1}$.   
The radial profile of the H-nuclei column densities, $\Nd$, is given in the right panel and is represented
by the thick solid curve joining the data points. 
The diamonds represent the column densities of H-nuclei in the molecular gas as inferred from CO $\Jone$ using a constant X-factor 
(see Section~\ref{Xfac}).  The squares represent the the column densities of H-nuclei in the atomic gas as inferred 
from HI 21-cm emission. For both the CO and HI, the error bars are comparable to or smaller than the symbols.  The dashed line gives the 
total gas column density as inferred from both CO and HI. 
}   
 \label{fig6a}
\end{figure*}
\clearpage


\begin{table}
 \caption{Gas Masses$\, ^a$ in \mfif\ and \meig.}
 \label{tab3}
 \begin{tabular}{@{}lll}
  \hline
   & \mfif & \meig \\
  \hline
  $M_d(H)\,^b$ & $9.2\times 10^9$ & $7.2\times 10^9$ \\
  $M(H_2)\,^c$ & $1.7\times 10^9\, ^d$ & $1.4\times 10^9\, ^e$ \\
  $M(HI)$ & $1.8\times 10^9$ & $2.2\times 10^9$ \\
  $M(H_2)+M(HI)$ & $3.5\times 10^9$ & $3.6\times 10^9$ \\
 \noalign{\smallskip}
  ${M(H_2)+M(HI)\over M_d(H)} $ & 0.4 & 0.5 \\
  \hline
 \end{tabular}

 \medskip

$^a$ All masses in M$_\odot$.

$^b$ Total gas mass as inferred from dust continuum emission.

$^c$ Molecular gas mass as inferred from the CO~$\Jone$ line.

$^d$ Using $\Xf = 0.8\Xtw$ for \mfif, see Section~\ref{Xfac}. 

$^e$ Using $\Xf = 1.0\Xtw$ for \meig, see Section~\ref{Xfac}. 


\end{table}










\subsection{X-Factor}\label{Xfac}

Maps of the X-factor, $\Xf$, are computed from the data, the details
of which are given in Appendix~\ref{appx}. 
The X-factor maps of Figures~\ref{fig13} and \ref{fig14} reveal spatial variations within the 
inner 7$\,$kpc radius of both galaxies.  The X-factor in \mfif\ is
on average larger in the interarm regions than in the arms by factors of roughly 2 or more.  This
is less obvious in \meig\ where the spatial resolution is lower (55$''$ due to the CO map), but is also
roughly the case.  

\begin{table}
 \caption{Mean$\, ^a$ X-Factor Estimates$\, ^b$ in \mfif\ and \meig}
 \label{tab5}
 \begin{tabular}{@{}llll}
  \hline
  \phantom{XXXX}&\mfif & \meig &\phantom{XXXX}\\
  \hline
  \phantom{XXXX}&0.8$^c$ & 1.0 &\phantom{XXXX} \\
  \hline
 \end{tabular}

 \medskip
 
$^a$ The $1/\sigma^2$-weighted means of the $\Xf$ map and only for the inner 7$\,$kpc radius.
     Uncertainty of about $\pm 50\%$.

$^b$ In units of $\Xtw$ or $10^{20} H_2\cdot cm^{-2}\cdot (\Kkms)$.

$^c$ Includes the correction determined from the simulations. 

\end{table}

The radial profiles  of $\Xf$ as seen in Figures~\ref{fig13a} and \ref{fig14a} 
show that $\Xf$ does not vary radially by more than a factor of 2 within the central 7$\,$kpc radius. 
In \mfif, such variations are consistent with a constant value for the inner 7$\,$kpc radius to within the
errors.  In contrast, $\Xf$ varies significantly within the central 3$\,$kpc radius of \meig: the ratio of the $\Xf$ 
at the bar ends to that in the central ``plateau'' is $0.50\pm 0.04$, significantly different from unity.   Even in 
the extreme low- and high-$\knu$ cases, these results are little changed. 

The X-factor maps of both \mfif\ and \meig\ hint at
spiral structure.  Applying the Fourier spiral analysis mentioned in Section~\ref{ssfa} to these X-factor maps yields a two-armed
spiral structure with a phase shift of roughly 90$^\circ$ with respect to the main spiral arms for both galaxies, meaning that   
the interarm/arm $\Xf$ ratios for \mfif\ and \meig\ are greater than unity with values 2.5 and 1.5, respectively.  These numbers apply 
only for the inner 8$\,$kpc and inner 6$\,$kpc radii for \mfif\ and \meig, respectively.  The uncertainties of these ratios 
is about 35\% for \mfif\ and 10\% for \meig.   
In the extreme low- and high-$\knu$ cases, the inferred interarm/arm ratios are 
the same to within the uncertainties.

The CO~$\Jone$ maps are noisy beyond galactocentric radii of about 7 to 8$\,$kpc.  Consequently, we are only able
to compute rough lower limits for $\Xf$ in the outer disks of \mfif\ and \meig. (See Appendix~\ref{appx} for details.)   
For \mfif, even in the high-$\knu$ (low-$\knu$) case, the lower limits to $\Xf$ are roughly $1$ to $30\Xtw$ 
($7$ to $10^3\Xtw$) depending on the radius in the outer disk.  For \meig, even in the high-$\knu$ (low-$\knu$) case, 
these rough lower limits to $\Xf$ are $0.3$ to $20\Xtw$ ($2$ to $600\Xtw$).  
{\it This is a 
strong hint that something unusual might be occurring in the gas or dust (or both) of the outer disks of \mfif\ and 
\meig\ regardless of the assumed value for the dust mass-absorption coeffient.\/}  See Section~\ref{xfsv} for 
details.

\begin{figure*}
\vspace{-30mm}
\includegraphics[height=150mm, keepaspectratio=true]{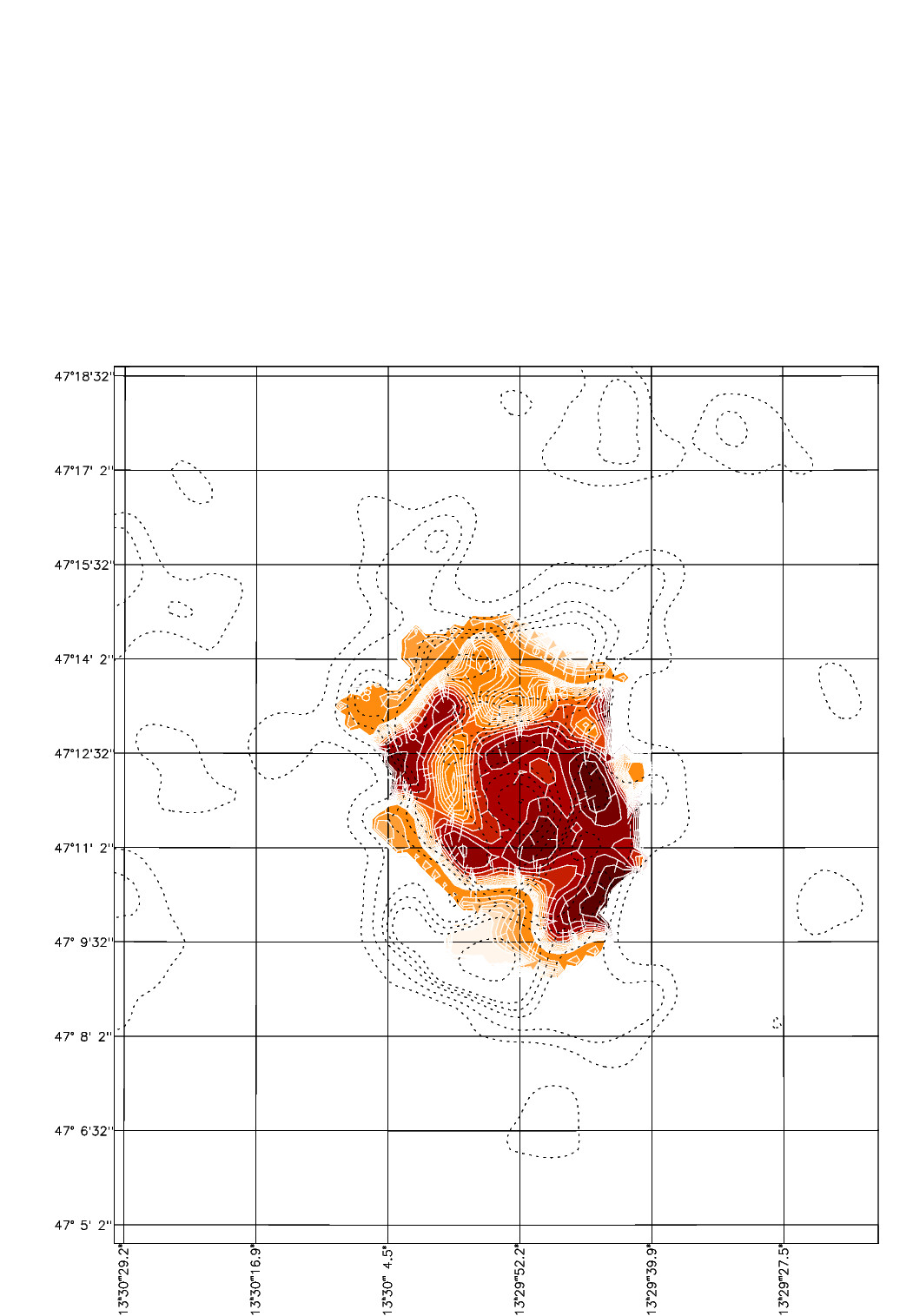}
\vspace{3mm}
 \caption{\mfif\ $\Xf$ map is shown with contour levels 0.2, 0.3, 0.4,..., 2.0, 4.0, 6.0,..., 14.0 $\Xtw$.  
 The dotted contours are the 1.1$\,$mm surface brightness with levels of 0.4, 0.6, 0.8, 0.9, 1.0, 1.4, 1.8,..., 3.0\mjysr.
}
\vspace{20mm}
\label{fig13}
\end{figure*}

\begin{figure*}
\vspace{-18mm}
\includegraphics[height=112mm,angle=90]{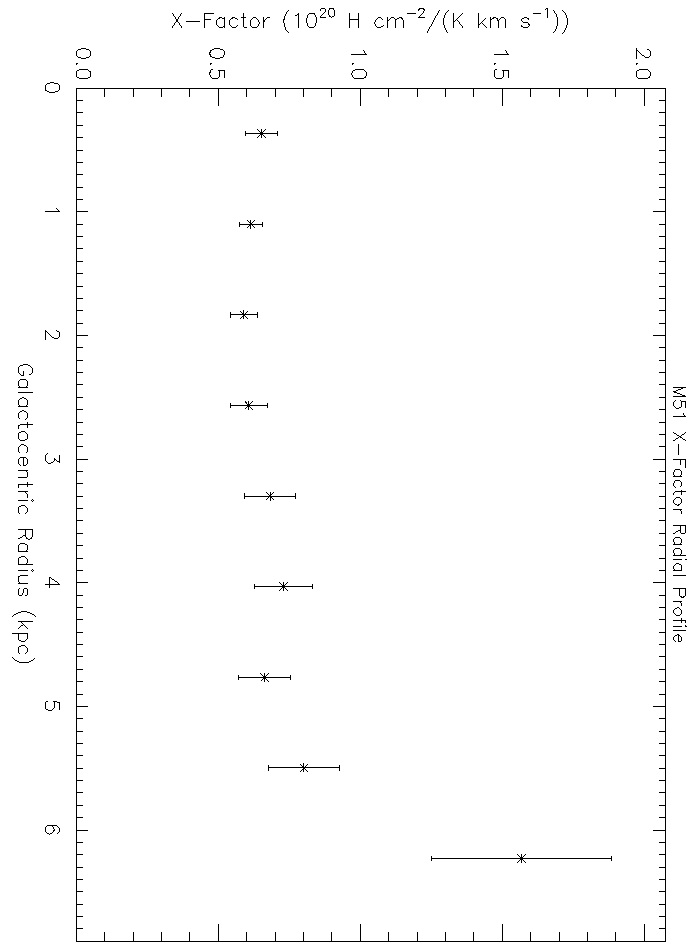}
\vspace{-1mm}
 \caption{The \mfif\ radial profile of the X-factor.  The azimuthally 
averaged X-factor is plotted against the galactocentric radius in kiloparsecs. The averages are determined within concentric annuli where 
each annulus is 2 pixels wide (18$''$ or 0.73$\,$kpc).  Note that the 23\% upward correction to the above X-factor values determined from the
simulations has not been applied. }   
 \label{fig13a}
\end{figure*}

\clearpage
\begin{figure*}
\vspace{-30mm}
\includegraphics[height=150mm, keepaspectratio=true]{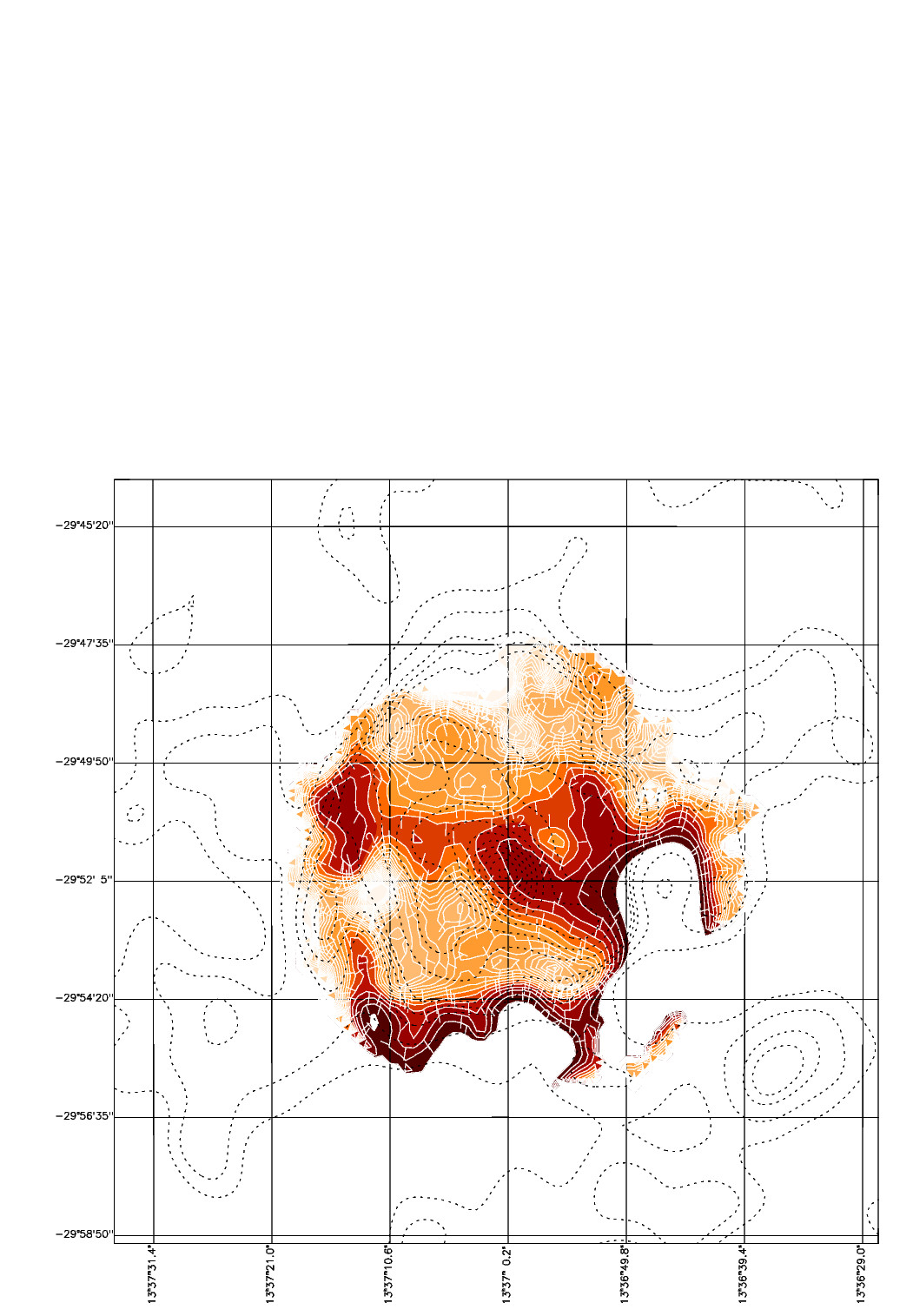}
\vspace{3mm}
 \caption{\meig\ $\Xf$ map is shown with contour levels 0.4, 0.6, 0.8,..., 5.2$\Xtw$.  
The dotted contours are the 1.1$\,$mm surface brightness with levels
of 0.4, 0.6, 0.8, 0.9, 1.0, 1.4, 1.8,..., 5.0\mjysr.
}
\vspace{20mm}
 \label{fig14}
\end{figure*}

\begin{figure*}
\vspace{-18mm}
\includegraphics[height=112mm,angle=90]{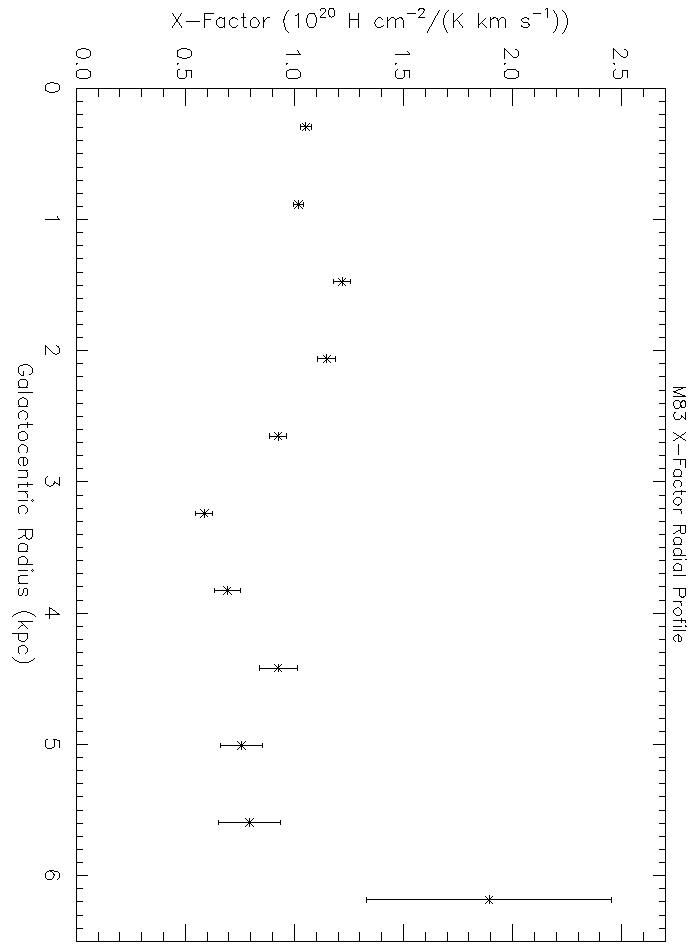}
\vspace{-1mm}
 \caption{The \meig\ radial profile of the X-factor.  The azimuthally 
averaged X-factor is plotted against the galactocentric radius in kiloparsecs. The averages are determined within concentric annuli where 
each annulus is 3 pixels wide (27$''$ or 0.59$\,$kpc).  
}   
 \label{fig14a}
\end{figure*}


\subsection{Star Formation versus Gas Surface Density}\label{sfvgsd}

\begin{figure*}
\vspace{0mm}
\includegraphics[height=210mm,angle=0]{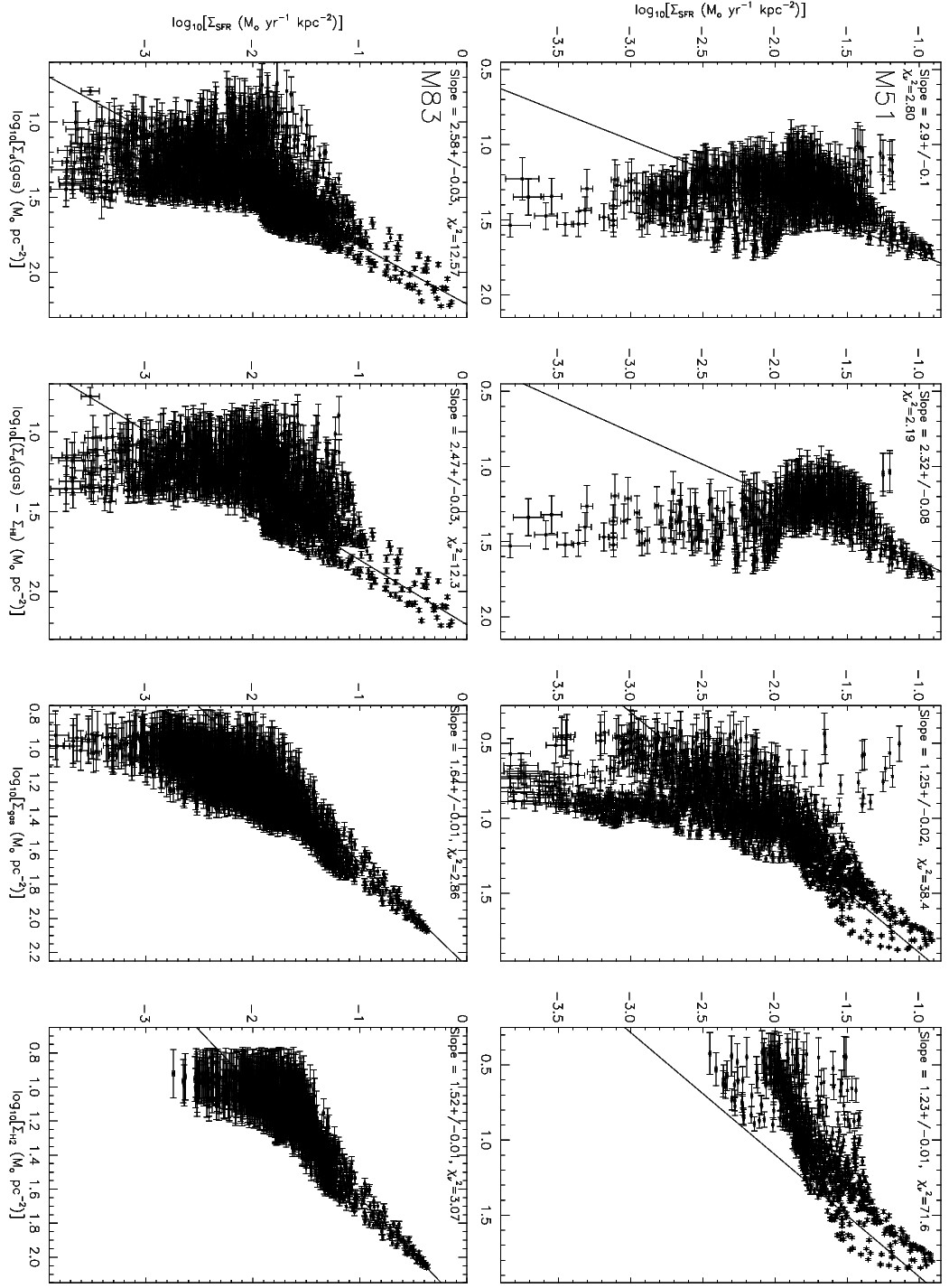}
\vspace{-0.2mm}
 \caption{The logarithm of the star formation rate (SFR) surface density is plotted against that from a tracer of gas surface density 
 in each of the four panels for \mfif\ (upper) and for \meig\ (lower).  In the first two panels for each row, the gas surface 
 densities are determined from the dust continuum.  In the last two panels for each row,  these surface densities are determined 
 from gas spectral lines only. The column densities in the first and third panels of each row are from tracers of the total gas 
 (molecular $+$ atomic) surface density and those in the second and fourth panels are from tracers of molecular 
 gas surface density only.  The linear fits are to the points where the $log_{10}[\Sfr(\msyrkp)]\geq -2.0$. The value of the 
 fitted slope and the reduced chi-square of the fit appear in the upper left of each panel.  
 }   
 \label{fig18}
\end{figure*}

Now we examine the variation of the star formation rate (SFR) surface density with the 
gas and dust surface density tracers.  (See Appendices~\ref{appsfrgas} and \ref{appsfrsurf} for details in
producing and comparing such surface density maps.)  
The plots of the surface densities of SFR versus gas are displayed in Figure~\ref{fig18}.
All that figure's panels, except the upper right, reveal an apparent gas surface density 
threshold of about 15$\mspc$ at which the SFR surface density rises nearly two 
orders of magnitude from $\sim 10^{-4}$ to $\sim 10^{-2}\msyrkp$.  Above this threshold, the 
SFR surface density follows a power-law rise with slopes of 2.5 to 2.9 for column 
densities inferred from continuum and with slopes of 1.2 to 1.6 for column densities inferred
from spectral lines. 

These superlinear slopes are consistent with the
inside-out star-formation scenario inferred for spiral galaxies by other means \citep[e.g.,][]{Gonzalez14}. 
In addition, the surface density threshold of $\sim 15\mspc$  visible in most of the panels is
roughly consistent with the threshold of $\sim 10\mspc$ determined from the simulations of \citet{Clark14}.
The simulations of \citet{Dobbs15} suggest a similar threshold in their Figure~11.

\citet{Bigiel08} also examined the star-formation law in external galaxies.   
Their fit using H$\alpha$ line and 24$\um$ data had a slope of 1.18.  This is comparable to the slope of
$1.23\pm 0.01$ for the corresponding plot of the current work.
The current work shows us that the slope changes yet
again -- to $2.47\pm 0.05$ -- when using continuum emission and the HI line as a tracer of the molecular gas, i.e, $\Sfr$ versus
[$\Sd - \Sh$].   A similar
difference in slopes between that for $\Sfr$ versus $\SH$ and that for $\Sfr$ versus $[\Sd - \Sh]$ is seen for \meig.
These varying numerical values of the slope and how they may or may not affect the physical interpretation of the relationship 
between the star formation and the gas will be discussed further in Section~\ref{sfrgsd}.
%

\begin{figure*}
\vspace{-22mm}
\includegraphics[height=200mm,keepaspectratio=true,angle=0]{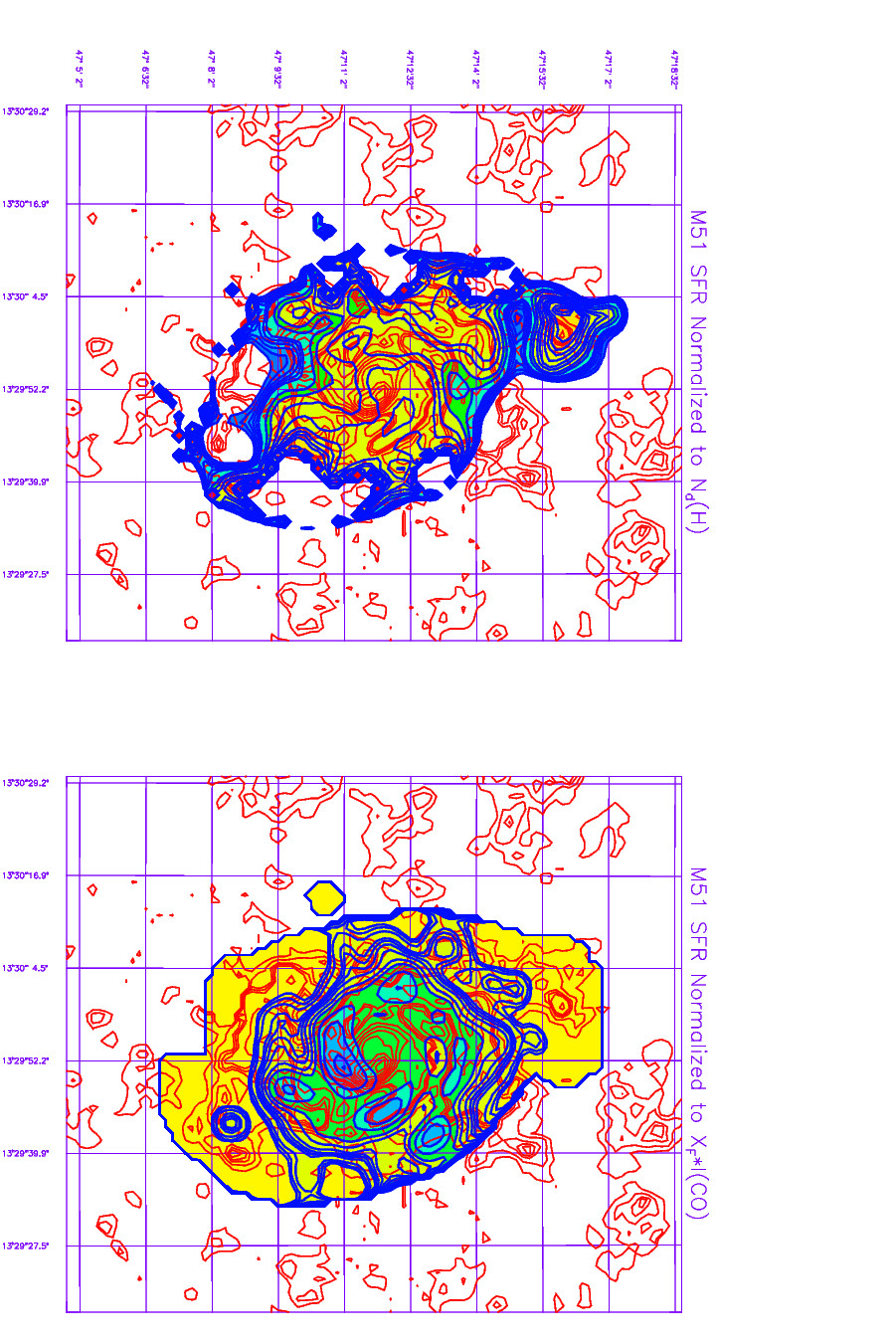}
\vspace{4.7mm}
 \caption{Maps of the SFR normalized to the molecular gas surface density --- sometimes called the star formation ``efficiency'' --- are
 shown for \mfif.   The left panel has the SFR normalized to the the difference between the continuum-derived gas surface density, $\Sd$, 
 and the surface density of atomic gas, $\Sh$.  The right panel is the SFR normalized to the molecular gas 
 surface density estimated from the adopted X-factor applied to the CO~$\Jone$ surface brightness.  The contour levels
 for the blue through green through yellow shaded areas for the left panel are 0.25, 0.30, 0.35, ..., 0.5, 0.75, 1.0, 1.5, 3.0, 4.0, 5.0, 
 6.25, 12.5, 18.75, 25.0$\, Gyr^{-1}$.  The contours of the right panel are those of the left panel scaled by 0.4.  The red contours in 
 both panels are the 1.1$\,$mm continuum surface brightnesses for \mfif\ with levels 0.4, 0.6, 0.8, 0.9, 1.0, 1.4, 1.8, 2.2, 2.6, 
 3.0, 3.4, 3.8$\mjsr$. 
 }   
 \label{fig19}
\end{figure*}

\begin{figure*}
\vspace{-6mm}
\includegraphics[height=200mm,keepaspectratio=true,angle=0]{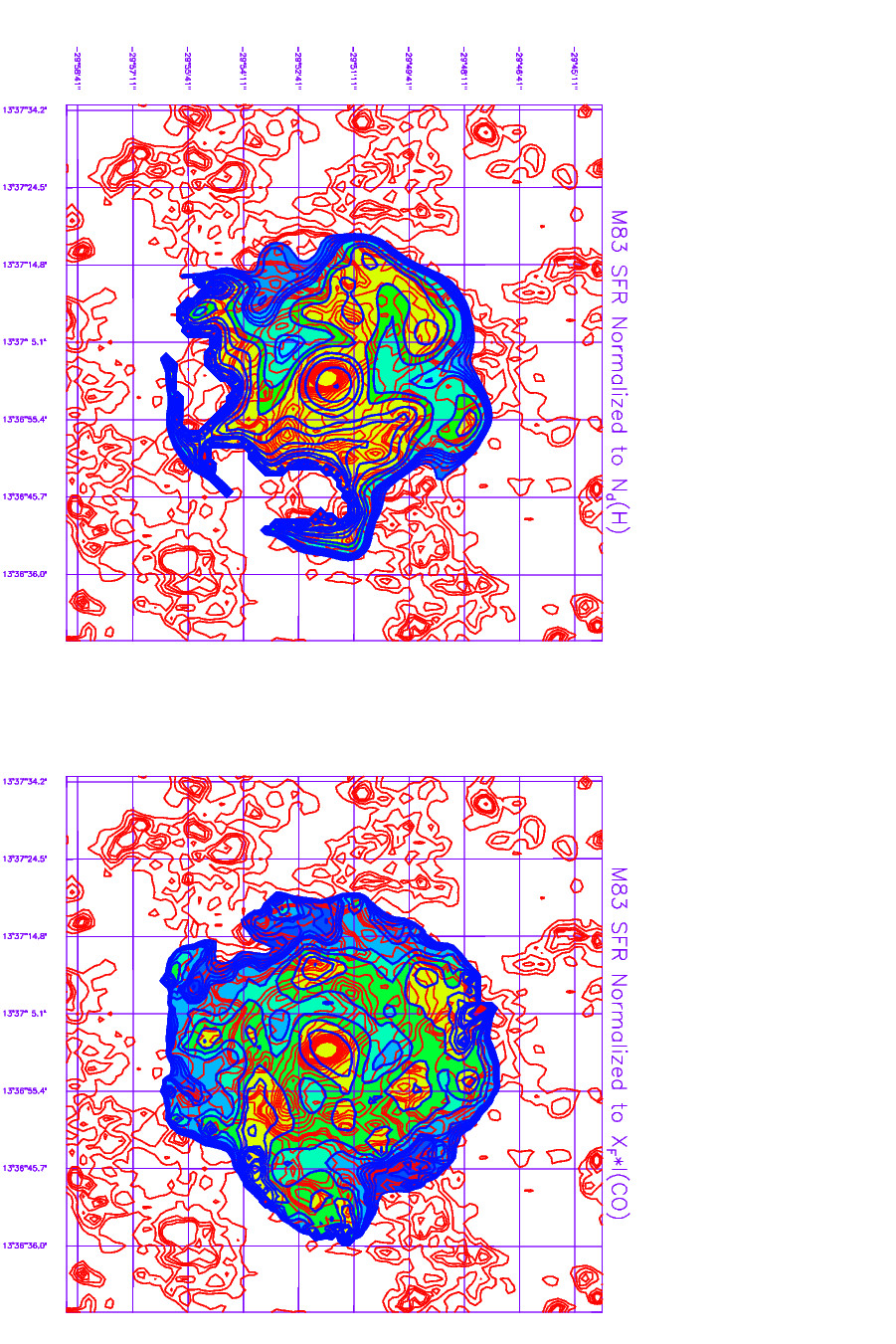}
\vspace{-1.2mm}
 \caption{Maps of the SFR normalized to the molecular gas surface density are
 shown for \meig.   The left panel has the SFR normalized to the 
 the difference between the continuum-derived gas surface density, $\Sd$, and the surface density of atomic gas, $\Sh$.
 The right panel is the SFR normalized to the molecular gas surface density estimated from the adopted X-factor 
 applied to the CO~$\Jone$ surface brightness.  The contour levels
 for the blue through green through yellow shaded areas for the left panel are 0.25, 0.30, 0.35, ..., 0.5, 0.75, 1.0, 1.5, 3.0, 4.0, 5.0, 
 6.25, 12.5, 18.75, 25.0$\, Gyr^{-1}$.  The contours of the right panels are those of the left panel scaled by 0.4.  The red contours in 
 both panels are the 1.1$\,$mm continuum surface brightnesses for \meig\ with levels 0.4, 0.6, 0.8, 0.9, 1.0, 1.4, 1.8, 2.2, 2.6, 3.0, 
 3.4, 3.8, 4.2, 4.6, 5.0, 5.4, 5.8, 6.2, 6.6, 7.0$\mjsr$.  
 }   
 \label{fig20}
\end{figure*}

More insights into star formation are provided by normalizing the SFR surface density to that of the molecular gas, producing
what is often called the star formation efficiency (SFE).   Images of the $\Sfr$ divided by molecular gas surface density
tracers, either $(\Sd - \Sh)$ or $2\,\Xf I(CO)$, are given in Figures~\ref{fig19} and \ref{fig20}.  
The yellow area surrounding much of the $\Sfr/\SH$ image of \mfif\ in 
the upper right panel suggests a very high star formation rate per unit gas mass in \mfif's outer disk.   But this is nothing more than 
an artefact due to using a constant and artificially low $\Xf$ rather than a higher, and likely more realistic, $\Xf$
for this outer area of the disk.  
The top image in Figure~5 of \citet{Foyle10} 
is that of their SFE of \mfif\ and the same artefact appears in the form of the white patches on the outer edges of the image.  (These
white patches also appear in the outer edges of the images of the other two galaxies, NGC$\,$628 and NGC$\,$6946, in that figure.) 


Whether or not spiral structure is visible in Figures~\ref{fig19} and \ref{fig20} is important for determining whether the 
spiral arms enhance the SFR beyond the corresponding enhancement in the gas surface density due to the arms.  Applying spiral 
arm decomposition to the images using
the dust-continuum derived surface densities yields arm/interarm ratios of 1.3 and 2.0 for 
\mfif\ and \meig, respectively.  
For the \mfif\ image using CO as the molecular gas tracer (right panel), the SFR normalized to the gas surface 
density is higher between the arms with interarm/arm ratio of 2.4.  For \meig, this is 1.0.  These ratios suggest some
mild effect on the SFR due to the spiral arms, but with the SFR normalized to surface densities from the continuum tracer suggesting an
opposite effect to that suggested by the SFR normalized to the surface densities from CO.   These apparently opposing effects are 
reconcilable when one considers the spatial variations of $\taun$ and of $\Xf$ as discussed in Section~\ref{sfrgsd}.

\clearpage

\clearpage

\clearpage 

\subsection{Spiral Structure Fourier Analysis}\label{ssfa}

Given that the large-scale structure of \mfif\ and \meig\ has been reliably recovered according to the simulations,
it is worthwhile to examine the spiral structure of both galaxies.  Hence, a two-dimensional Fourier analysis on the
basis of logarithmic spirals \citep{Kalnajs75, Considere88, Puerari92, Block99} was conducted on different images of
both \mfif\ and \meig, thereby allowing tests of the spiral structure.  

\begin{figure}
\vspace{-2mm}
\includegraphics[height=50mm,angle=0]{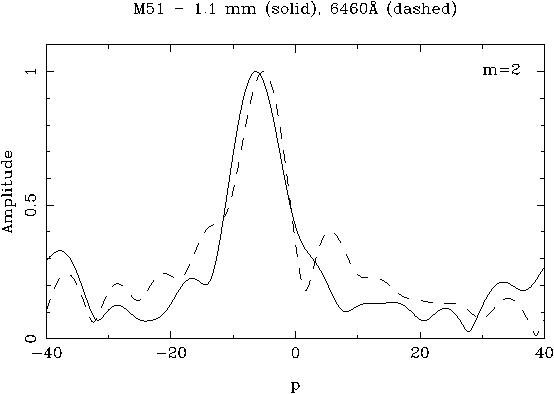}
\vspace{-2mm}
 \caption{Fourier transform results for the logarithmic spiral in the 1.1$\,$mm and R-band images of \mfif. 
The power spectrum for the $m=2$ spiral is shown for both images as a function of $p$, which is a measure of
the pitch angle (i.e., the $p$ is related to pitch angle, $P$, by $p= -m \cot(P)$).   The main peak of each 
spectrum gives very similar pitch angles for both the 1.1$\,$mm ($P=18^\circ$) and R-band ($P=20^\circ$)
images .
}   
 \label{fig20.1}
\end{figure}

\begin{figure}
\vspace{-2mm}
\includegraphics[height=50mm,angle=0]{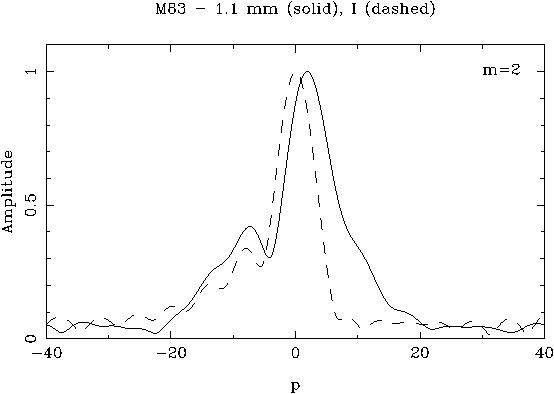}
\vspace{-2mm}
 \caption{Similar to Figure~\ref{fig20.1}, Fourier transform results for the logarithmic spiral in the 1.1$\,$mm 
and $I$-band images of \meig.   The power spectrum for the $m=2$ spiral is shown for both images as a 
function of $p$.   For this galaxy, in the radial range being analyzed, the $I$ image is dominated by the 
oval/bar distortion.
This power-spectrum shows a clear peak at 
$p=0$ ($P=90^\circ$).
}   
 \label{fig20.3}
\end{figure}

One obvious test is to see whether the spiral structure observed in visible (or nearly visible) 
light,
due to stars, is the same as that 
observed in the millimetre continuum, due to dust (and its associated gas).  Accordingly, the spiral-arm analysis
mentioned above is applied to an R-band image of 
\mfif\footnote{See \hbox{$http://ned.ipac.caltech.edu/cgi$-$bin/ex\_refcode?refcode=1994DSS...1...0000\%3A$ for \mfif.}} 
and an I-band image of 
\meig\footnote{See \hbox{$http://ned.ipac.caltech.edu/cgi$-$bin/ex\_refcode?refcode=2000ApJS..131..441K$ for \meig.}}, 
as well as the
{\it AzTEC\/} 1.1-mm image. 
Figure~\ref{fig20.1}, for example, shows that the power spectrum in \mfif\ for 
the $m=2$ (i.e. two-arm) spiral pattern is identical, to within the uncertainties, between the R-band and the 1.1$\,$mm 
continuum \mfif.  However, the power spectrum in Figure~\ref{fig20.3}
for \meig\ reveals a noticeable difference between the spiral structure in the 1.1$\,$mm image and that in the $I$-band
image; only a hint of a small bar is discernible in the former image, whereas a more prominent bar adorns the latter image. 
Further analysis finds that the visible (or near visible) light and 1$\,$mm continuum spirals are not offset from each other.   
Accordingly, this suggests that the stars and gas are also not offset from each other in the spiral arms. 

\begin{figure*}
\vspace{-2mm}
\includegraphics[height=90mm,width=155mm,angle=0]{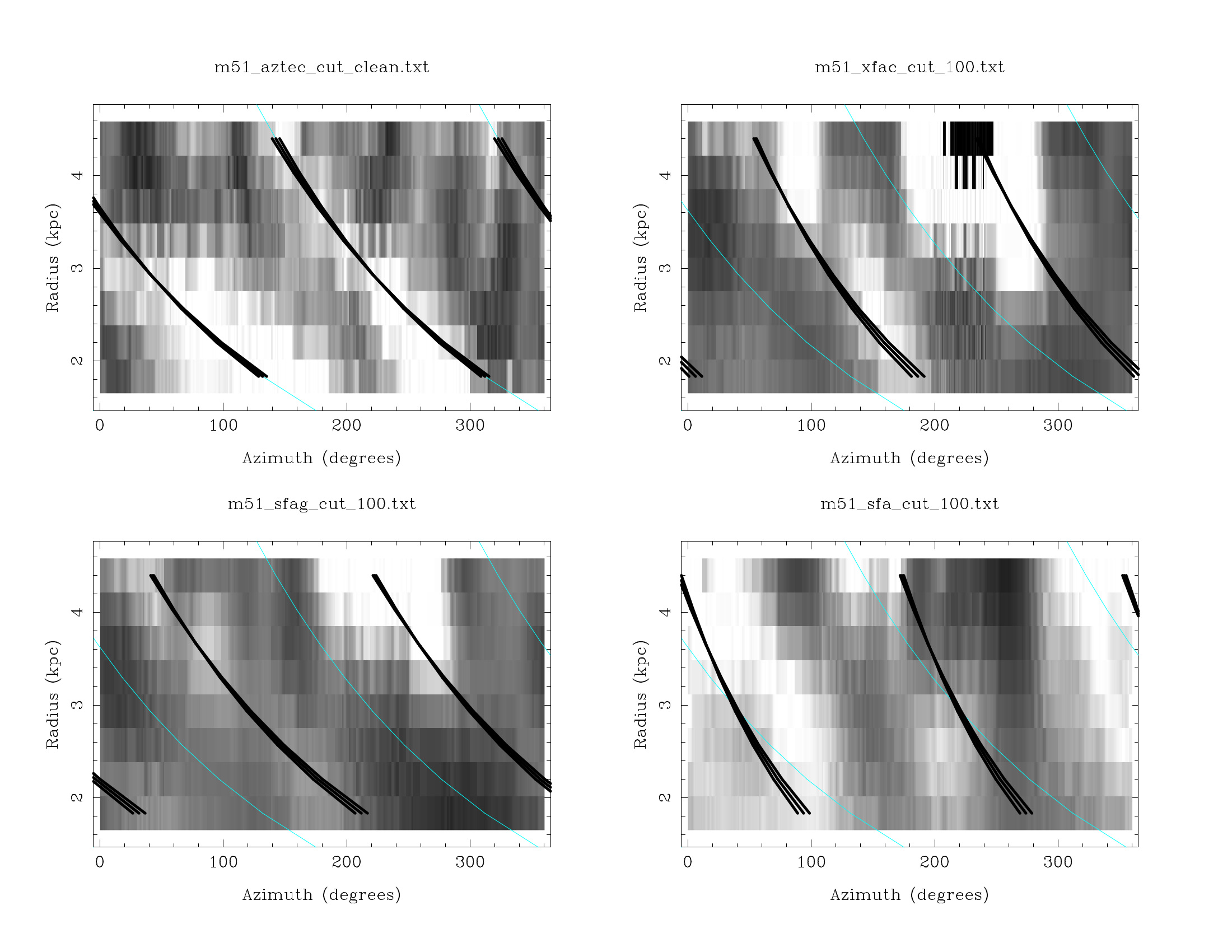}
\vspace{-4mm}
 \caption{Fourier transform results overplotted on the radius versus theta images for \mfif. Upper left: {\it AzTEC\/} image; 
upper right, X factor; bottom left, SFAG image; bottom right SFA image. The radius in each panel is in kiloparsecs. Theta is in 
degrees, where zero is for the west and increases anticlockwise. 
The minimum to maximum radii in which we conduct the 2D Fourier analysis are 1.8 to 4.4$\,$kpc. 
The thin blue lines in the four panels represent the spiral structure 
we detect in the {\it AzTEC\/} image (pitch angle of 18 degrees). The thick black lines represent the bisymmetrical structure 
we detect on all images. As clearly seen, the spirals of the X factor and SFAG images (both with pitch angle of 
17 degrees) are rotated with respect to the {\it AzTEC\/} image. This means that the the X factor and SFAG images have 
higher values in the interarm regions. The spiral structure we detect on the SFA image has a smaller pitch angle (around 13 
degrees), but it is in phase with the {\it AzTEC\/} arms.
 }   
 \label{fig20a}
\end{figure*}

\begin{figure*}
\vspace{-4mm}
\includegraphics[height=90mm,width=145mm,angle=0]{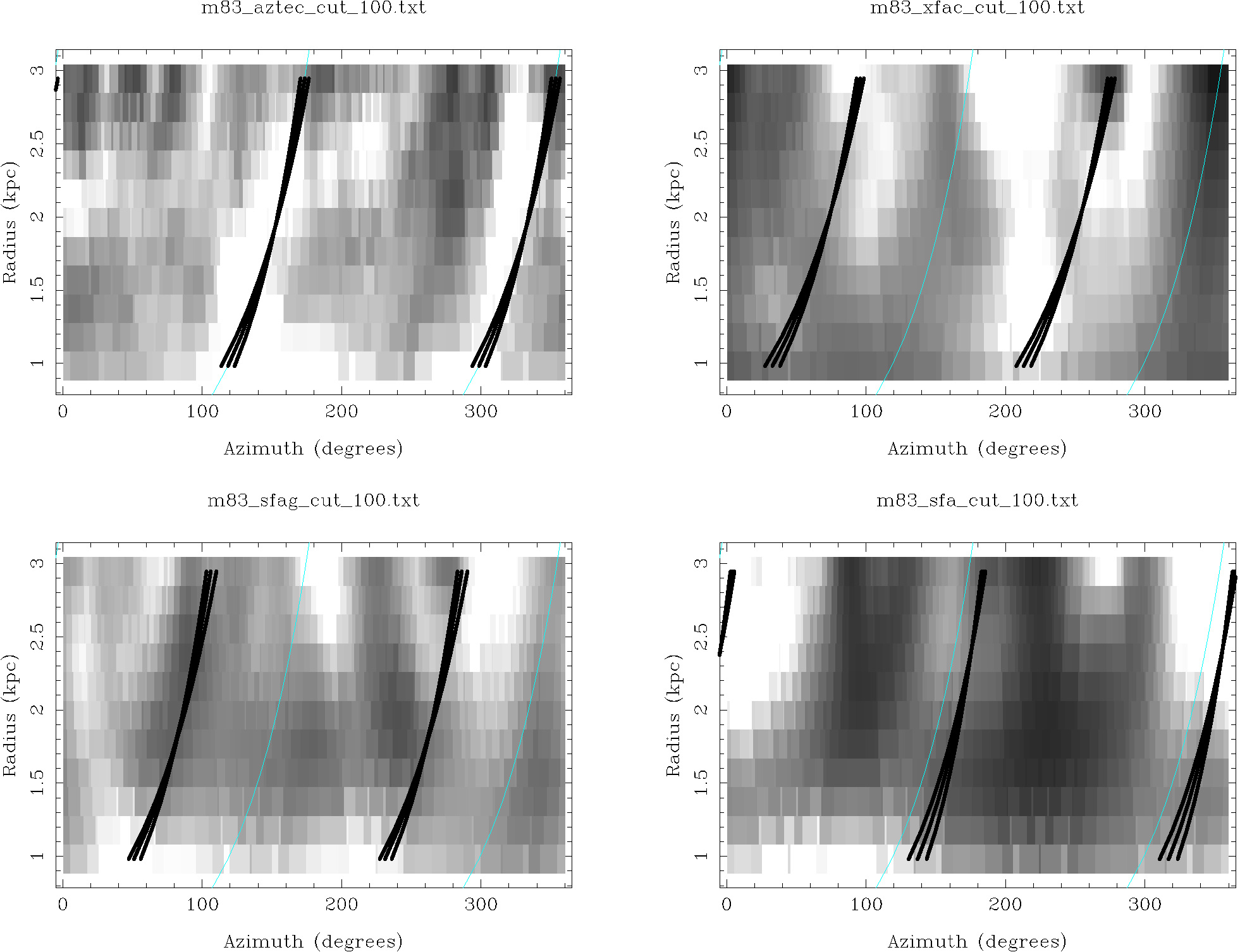}
\vspace{-1mm}
 \caption{Fourier transform results overplotted on the radius versus theta images for \meig. Upper left: {\it AzTEC\/} image; 
upper right, X factor; bottom left, SFAG image; bottom right SFA image. The radius in each panel is in kiloparsecs. Theta is in 
degrees, where zero is for the west and increases anticlockwise. 
The minimum to maximum radii in which the 2D Fourier analysis is carried out are 1.0 to 3.0$\,$kpc. 
The thin blue lines in the four panels represent the spiral structure 
we detect in the {\it AzTEC\/} image (pitch angle of 48 degrees). The thick black lines represent the bisymmetrical structure 
we detect on all images. The spirals of the X factor has a pitch angle of 31 degrees, which is similar to that of 36-degree
pitch angle of the SFAG image.   The pitch angle of the SFA image is 53 degrees.  As found for \mfif, the X-factor and SFAG 
spirals are rotated with respect to the {\it AzTEC\/} image, again yielding higher values in the interarm regions.  
}   
 \label{fig20b}
\end{figure*}

Other tests of spiral structure are applied to the X-factor and to the star formation rate surface density normalized
to the gas surface density, $\Sfr/\Sg$ or $\Sfr/\SH$.  The results of these tests are presented in Figures~\ref{fig20a} 
and \ref{fig20b} for \mfif\ and \meig, respectively, which display the results of this 
Fourier analysis for the 1.1$\,$mm continuum images, the X-factor maps, and the maps of the
SFR normalized to the molecular gas surface density, where that surface density is determined 
from the continuum and HI for the ``SFA'' panel and determined from CO for the ``SFAG'' panel. 
Clearly the higher values for both the X factor, and the SFAG images are found in the interarm region, and the spiral 
structure have almost the same pitch angle as the main {\it AzTEC\/} arms. For the SFA images, the pitch angle is a bit 
smaller, but it is in phase with the arms we detect in the {\it AzTEC\/} image. These results are similar between 
\mfif\ and \meig.

We have used the detected positions of the arms in the {\it AzTEC\/} images of \mfif\ and \meig, and calculated the arm-to-interarm 
ratios of the processed images.  These are presented in Table~\ref{tab7}.  The arm-to-interarm ratios are for the inner
disks of both galaxies --- galactocentric radii of 1.8-5.5$\,$kpc and 1.0-2.9$\,$kpc for \mfif\ and \meig, respectively.  
These radii were chosen for consistency with the X-factor maps.  
Most of the numerical values listed in Table~\ref{tab7} are within a factor of 2 of
unity. The spiral arms seen in the 1.1$\,$mm continuum have ratio values that are comparable to those seen in red light or $I$-band.
The arm/interarm values of the X-factor images indicate that the X-factor is higher in the interarm regions, where those
interarm regions are defined as those in the millimetre continuum and in the $I$ and $R$ bands (i.e., between the dust and stellar arms).


Similar to the X-factor, the SFAG map is higher between the arms than in the
arms, at least for \mfif.  
Given that the SFAG map was computed from the X-factor, it is not surprising that both X-factor and SFAG maps have this quirk in 
the arm-to-interarm ratio.  
Compensating the arm/interarm ratio of SFAG for that of the X-factor suggests that the ``true'' arm-to-interarm ratio is above unity.  
Indeed, that is confirmed in the SFA map, which is the SFR surface brightness normalized to the molecular gas surface
density determined from the continuum and the HI line and, as a result, the SFA map is independent of the X-factor.   

Taken at face value, this suggests that star formation is more ``efficient'' in the arms than in the interarms, in the sense that the 
star formation rate in the arms is enhanced beyond that expected from simply having more gas and dust surface density in the arms.
However, the values of the arm-to-interarm ratios presented here are subject to systematic effects.  See Section~\ref{sfrgsd} for
more discussion of this. In any event, even if real, this enhancement is less than a factor of 2 or even 1.5. 

\clearpage

\clearpage

\begin{table}
 \caption{Spiral Arm-to-Interarm Ratios in the Inner Disks$^a$ of \mfif\ and \meig}
 \label{tab7}
 \begin{tabular}{@{}lll}
  \hline
  Image & \mfif & \meig \\
  \hline
  Visible$^b$ & 1.3  & 1.5  \\
  1.1$\,$mm    & 1.7  & 1.4  \\
  X-Factor     & 0.4  & 0.7 \\
  SFA$^c$      & 1.3  & 2.0  \\
  SFAG$^d$     & 0.4 &  1.0   \\
  \hline
 \end{tabular}

 \medskip
 
$^a$ For galactocentric radii of 1.8-5.5$\,$kpc and 1.0-2.9$\,$kpc for \mfif\ and \meig, respectively.

$^b$ In red light for \mfif\ and $I$-band for \meig.

$^c$ The star formation surface density map normalized to the molecular gas surface density determined 
     from the continuum and the HI line.
     
$^d$ The star formation surface density map normalized to the molecular gas surface density determined 
     from the CO $\Jone$ line and the X-factor.

\end{table}

\section{Discussion}

Maps of \mfif\ and \meig\ were made in the 1.1$\,$mm continuum from observations with the instrument {\it AzTEC\/} with the 
{\it JCMT\/}.   Combining with these maps with the corresponding {\it Spitzer\/} 160$\um$ (or, more properly, 155.9$\um$) 
maps gave estimates of the gas surface densities in these two galaxies (see Appendix~\ref{appsurf} and \ref{appgsdm} for a 
detailed discussion).  With these gas surface density maps, spatial variations of the X-factor were estimated.  In 
addition, we investigated the relationship between the gas surface density and that of the star formation rate.   These 
are dealt with in more detail below.


%
%
%
%
%
%
%

\subsection{The X-Factor, its Spatial Variations, and CO-Dark Gas}\label{xfsv}

The most important results of this work regarding the X-factor are the following:
\begin{enumerate}
 \item The average X-factor for each galaxy can be estimated from the current 
       observations, even if crudely.  Those average values are $\sim 0.8$ and 
       $\sim 1.0\Xtw$ for \mfif\ and \meig, respectively.
 \item The X-factor is higher in the interarm regions than in the arms.
 \item There seems to be CO-dark gas that resides mostly in the outer disks of
       both \mfif\ and \meig.
\end{enumerate}
The latter two results are robust to a range of adopted $\taun$ values. 

The variation of the X-factor spatially and from source to source could be due, in part, to
variations in metallicity. 
Theoretical work using the observational data also support a dependence of $\Xf$ 
on metallicity \citep[e.g.,][]{Nar12, Lagos12}. In contrast, \citet{Sandstrom13} do not find
a strong correlation of $\Xf$ with metallicity.  
However, their sample only had a metallicity range of 0.5-0.8 dex within factors of 
~3 of solar.   The irregular galaxies observed by Israel and
others \citep[e.g., see][]{Israel88, Dettmar89, Israel97, Israel97a, Madden97, 
Fukui99} typically had metallicities much less than solar, sometimes only a few percent
of solar, and found X-factors an order-of-magnitude or more higher than the standard
value.  So a strong $\Xf$-metallicity relation may exist for galaxies with 
strongly sub-solar metallicities.  For galaxies with roughly solar metallicities, 
while metallicity alone is apparently insufficient in constraining $\Xf$, it is still relevant.   
For example, using the data points for NGC4321 from the left panel of Figure~10 in \citet{Sandstrom13} 
yields a correlation coeffient of $-$0.6.  This suggests that each galaxy has its own $\Xf$-metallicity
relationship. 

Given that sub-solar metallicities imply larger X-factors, perhaps the interarm regions of
\mfif\ and \meig\ have sub-solar metallicities, while being at solar-level in their arms 
\footnote{\citet{Parkin13}, however, find that the FUV flux and molecular gas density are
the same for arm and interarm regions in \mfif.  Whether this implies the same metallicity
in both regions is unclear.}.   The models of \citet{Nar12}, as well as their Figure~1, suggest 
that the metallicity in the interarm regions would be systematically lower by a factor of
$\sim$2-3 in order to increase $\Xf$ by a factor of 2 with respect to that in the spiral arms
(see their equation~\#8).  While observations of metallicity in the ISM of \mfif\ and
\meig\ apparently do not support such a systemically lower metallicity between their spiral
arms \citep[see][]{Bresolin04, Bresolin09}, they also do not rule it out: such observations
are toward HII regions and are heavily biased toward the spiral arms. 

This could, in turn, affect the results of studies of the effects of
the spiral arms on star formation.  
\citet{Foyle10}, for example, looked at whether the SFR normalized to
the molecular gas surface density is higher in the spiral arms of three galaxies.  They
adopted a spatially constant X-factor, which, to within a factor of 2, is likely correct.
This will be discussed further in Section~\ref{sfrgsd}.

The modelling by \citet{Nar12} mentioned above suggests that the X-factor depends mainly 
on the metallicity of the gas in a galaxy and, to a lesser extent, on the average CO surface brightness. 
Using the roughly solar metallicities of \mfif\ and \meig\ \citep[see][]{Bresolin04, Bresolin09},
the observed CO brightnesses from the data used here, and applying expression~(8) of \citet{Nar12}
yields $\Xf\simeq 3\Xtw$ for \mfif\ and $\Xf\simeq 3$ to 6$\Xtw$ for \meig; this is much higher than 
the observed values found in the current work.
Admittedly, our estimates are uncertain by factors of about 2.   Nevertheless, our 
estimate of $\Xf\simeq 0.8\Xtw$ for \mfif\ agrees with the result of \citet{Nakai95} 
who find $\Xf = 0.9\pm 0.1\Xtw$ from using the observed extinction in HII regions. 
Accordingly, the theoretical models  need further adjustments.


As mentioned previously, there is evidence that adopting a spatially constant X-factor 
does not account for CO-dark gas.  
For example, 
Table~\ref{tab3} suggests that about half the total gas mass is 
unaccounted for when using the CO~$\Jone$ and HI~21-cm spectral lines (see Section~\ref{Xfac}
and Appendix~\ref{appgsdm}).  If we adopt a 
threshold for CO-dark gas that corresponds to an $\Xf$ that is a factor of 4 higher than 
the average for each galaxy, then some of the edges of the shaded regions of 
Figures~\ref{fig13} and \ref{fig14} indicate such gas.  It is worth noting that
these edges are well within the boundaries of the CO maps.

Of course, possible alternative interpretations for the high X-factor or its high lower limits 
are not entirely ruled out.  These include the following:
\begin{itemize}
 \item An extended low surface brightness artifact in the 1.1$\,$mm continuum maps of \mfif\ and \meig. 
 \item Insufficient mapping of CO~$\Jone$ in the outer disks of these galaxies.
 \item Dust with unusual properties such as unusually high dust-mass absorption coefficient
       (i.e., $\knu$) and/or a high dust-to-gas mass ratio.
 \item Optically thick HI 21-cm emission.
\end{itemize}
Each of the above could mimic the presence of CO-dark gas.  The first alternative is unlikely
given that the simulations of the {\it AzTEC\/} 1.1$\,$mm observations have accurately accounted
for any constant offsets in the \mfif\ and \meig\ maps.  
The second alternative is unlikely because 
there is evidence for such CO-dark 
gas seen at radii well within the boundaries of the existing CO maps, as mentioned previously.  
The third alternative is unlikely because 
having a combination of high $\knu$ and high $x_d$ would not be sufficient for positions 
with $\Xf$ two orders of magnitude larger.  

The fourth alternative is a partly valid explanation for CO-dark gas in our Galaxy according to the 
\citet{Parade11a}.  They estimated that up to half of the dark gas could be due to optically thick 
HI 21-cm emission. Even if that were the case for \mfif\ and \meig, it would not account for lower 
limits to $\Xf$ that are one or two orders of magnitude higher than the average inner disk value. 

In short, none of the alternatives mentioned above are likely to entirely rule out CO-dark gas.  
Nevertheless, these alternatives themselves are {\it not\/} entirely ruled out either and could 
partly account for some of the high X-factor values inferred.  In any event,
{\it more and deeper mapping of CO and other gas tracers of the outer disks of these galaxies is essential for 
understanding the nature of the dust and gas at these large galactocentric radii.\/}

The existence of CO-dark gas in our Galaxy has been known for a while \citep[see][and references therein]{Reach98}
and has been confirmed recently by the \citet{Parade11a}.  They find $\Xf = 2.5\pm 0.1\Xtw$ for our Galaxy.  Also, as
mentioned previously there is much additional evidence for such gas in the Galaxy and external galaxies 
\citep[see, e.g.,][]{Israel97, Israel97a, Baes14, Clark12, Langer14, Pineda14, Roman10, Saintonge12, Smith14}.
So it is quite likely that there is CO-dark gas in both \mfif\ and \meig.  Indeed, the current observations 
suggest that such gas is present in the outer disks of both \mfif\ and \meig.  The CO-dark
gas could account for as much as half of the total gas mass, although the alternative interpretations given above
could reduce that fraction by roughly 50\%.



\subsection{The Relationship between the Star Formation Rate and the Gas Surface Density}\label{sfrgsd}


In Section~\ref{sfvgsd}, we examined how the SFR varied with the gas surface density in a number of ways. 
Comparing between the continuum and spectral line tracers given in Figure~\ref{fig18}, 
all show higher slopes for the continuum tracers than for the spectral-line tracers.  
Figures~\ref{fig5a} and \ref{fig6a} illustrate that the continuum-derived radial profile of 
the gas surface density is flatter than that for the spectral-line derived gas surface for large galactocentric 
radii.  Having a smaller range of gas surface densities will naturally increase the slope in the SFR versus gas 
surface density plots.


Even though \citet{Bigiel08} used far-UV and 24$\um$ data 
to estimate the SFR, whereas we used H$\alpha$ and 24$\um$, that is unlikely to account for the difference between their
fitted slopes and ours; their Figure~9 makes such a comparison and the difference in fitted slopes between the two
SFR tracers is 10\% or less.  So if differences in SFR tracers cannot explain the difference between the slope obtained
by \citet{Bigiel08} and those obtained in the current work, then it must be the difference in gas surface density tracer.
The continuum tracer used here gives slopes of $\sim 2.5$, which is consistent with the eyeball inspection
of the far upper right panel of Figure~8 of \citet{Foyle12} for the inner $\sim 6\, kpc$ radius of \meig. 

Estimation of gas surface or column densities is problematic because, as exemplified in the current work,
different tracers can yield different results.  There are systematics that affect the column density estimates using
CO~$\Jone$ and different systematics when using infrared and millimetre continuum.  These systematics yield
the differing slopes and also the observed scatter, which is sometimes quite large with {\it reduced\/} chi-square 
as high as $\sim 40$ to 70.  This scatter is partly intrinsic, because the SFR depends on more 
physical conditions than on just column density.
But systematic errors in estimating surface densities also play a role.  
Corrections applied systematically to those errors would represent a smooth gradient with 
galactocentric radius and not simply corrections to a few individual points.  For example, observations have 
inferred X-factor values that are factors of about 5 or more lower than the standard 
value \citep[e.g.,][]{Rickard85, Israel88, W93, Regan00, Pag01} in the centres of external 
galaxies as well as in the central region of our own Galaxy \citep{Sodroski95, Dahmen97, Dahmen98}.  But these 
centres only represent the central few hundred parsec radii and would thus represent very few points in the plots
of Figure~\ref{fig18}.  Only a large, smooth gradient in parameters like the X-factor or $\taun$ would change the 
slope uniformly in those plots, and there is no evidence for such. 

Nevertheless, there is evidence of a weak dependence of the X-factor on the SFR.   \citet{Clark15} explored
this possibility with their physical models of molecular clouds, finding that $\Xf\propto\Sfr^\gamma$ with 
$\gamma\simeq 0.5$.  The systematically different power-law indices for the dust tracers from those for
the gas tracers are reconcilable by adopting that relationship between $\Xf$ and $\Sfr$.   The 
computation of Appendix~\ref{appxfsfr} yields $\gamma = -0.38$ and $-0.25$ for \mfif\ and \meig, respectively.
These have uncertainties of 3-4\% and are thus significantly different from that of the models of \citet{Clark15}.    
These are more consistent with the models of \citet{Nar12}\citep[see][]{Clark15}.   That the \mfif\ and
\meig\ $\gamma$ values are significantly different from each other argues that there is no universal relationship
between $\Xf$ and $\Sfr$.  This is not surprising given that both of those quantities have complex dependences
on the physical structure within the molecular gas. 

The value of the slope of the surface densities of SFR versus those of the molecular gas is an indication of the 
large-scale evolution of a galaxy.  As briefly alluded to previously, following the ``stream'' of 
points in each of the panels of Figure~\ref{fig18} from low to high surface densities is equivalent 
to travelling from large to small galactocentric radii.  
The $\Sfr/\SH$ (or
alternatively $\Sfr/[\Sd - \Sh]$) ratio is really the inverse
depletion time of the molecular gas.  For slope $=1$, $\Sfr/\SH = constant$ and the gas is depleted uniformly throughout
a galaxy, as pointed out by \citet{Bigiel08} and found by them and \citet{Leroy13} .   However, 
{\it all\/} the panels of Figure~\ref{fig18} have slopes $>1$, 
strongly supporting an inside-out depletion of molecular gas in both \mfif\ and \meig.  

The panels of Figures~\ref{fig19} and \ref{fig20} yield molecular gas depletion times from
their centres out to radii of about 8$\,$kpc for \mfif\ and 6$\,$kpc for \meig.   For \mfif, the continuum tracers 
give molecular gas depletion times of about 1.2$\,$Gyr in the centre to 20$\,$Gyr in the outer disk and, with the 
CO line tracer, these times are 0.8 to 2.5$\,$Gyr.  For \meig, the continuum tracers suggest depletion times 
of about 0.7$\,$Gyr in the centre to 10$\,$Gyr in the outer disk and, again for CO, these times are about 0.4 to 
5$\,$Gyr.   This inside-out evolution of the star formation in the disks of galaxies is supported by the visible-light
observations of  \citet{Gonzalez14}.  

At radii from about 8$\,$kpc to 11$\,$kpc for \mfif, 
the molecular
gas depletion times at these radii extend by nearly two orders of magnitude higher than the 20$\,$Gyr estimate for
galactocentric radius of 8$\,$kpc.  For \meig, this extension of the gas depletion time occurs for radii slightly
beyond 6$\,$kpc and is by about 1.5 orders of magnitude.

Figures~\ref{fig19} and \ref{fig20} are equivalent to Figure~5 of \citet{Foyle10}, displaying the inverse
depletion times of the molecular gas in the form of images of entire galaxies.  Those images in the current paper 
\citep[and, to some extent, those of][]{Foyle10} are suggestive of spiral structure.   Such spiral structure
implies an enhancement of the SFR due to the spiral arms beyond that of arms simply collecting and compressing gas and dust.  
Accordingly, we subjected these images to spiral arm Fourier analysis and found only a weak
spiral structure with arm/interarm ratios usually within factors of 2 of unity.    The right panels were found to 
have a spiral structure with high {\it interarm\/} values of $\Sfr/\SH$ and lower arm values, which is corrected when
accounting for the higher interarm $\Xf$ (see Section~\ref{Xfac}).
In contrast, the left panels were found to have a spiral structure with high $\Sfr/[\Sd -\Sh]$ on the arms and lower values 
between the arms; the arm/interarm ratios are 1.3 for \mfif\ and 2.0 for \meig.  These arm/interarm ratios are at least 
partly explained by the higher $\taun$ between the arms than in the arms (see Appendix~\ref{appgsdm})--- 18\% higher
for \mfif\ and 8\% higher for \meig.  

So, conservatively speaking, even the continuum tracer in our work {\it confirms the work of \citet{Foyle10} that 
spiral arms only enhance the star formation rate because of increasing the surface density of gas and dust with\/} 
no {\it additional enhancement.\/}  The uncertainties in the current work do not permit completely ruling out such 
an enhancement, but suggest that any such enhancement would be small (i.e., a factor of $\lsim 2$).   That enhancement,
should it be real, could be accounted for by orbit-crowding in the spiral arms raising the inverse depletion time of 
the gas in those arms \citep[e.g.][]{Moore12}.

One key question is whether the conclusions are still valid if the diffuse emission is removed \citep[][]{Foyle13}.
But the lower spatial resolution of the gas tracer observations impede determination of the surface densities associated
only with the star-forming regions.  So a surface-density versus surface-density plot is difficult to create 
\citep[see][]{Foyle13}, making comparison in the context of much previous work difficult.
Removing the diffuse emission {\it might\/} still result in the slope of the $log (SFR\ surface\ density)$ versus 
$log (gas\ surface\ density)$ still being higher when a dust-continuum tracer is used in place of a gas-line tracer 
(see Figure~\ref{fig18}), due to the CO-dark gas not traced by spectral lines or maybe due to $\Xf$ varying with the SFR.
Also, the inside-out galactic-scale evolution of
star formation, as indicated by the superlinear slopes, is likely still valid due to the support of independent work
\citep{Gonzalez14}. 

\citet[][]{Shetty13} use hierarchical Bayesian linear regression on the observational data, finding that no one S-K relation
holds for all galaxies. This is consistent with the current work where no one power-law applies to either \mfif\ or
\meig.  Indeed, any simple power-law fit is inapplicable given the poor quality of fits in Figure~\ref{fig18}.
Not only are there different slopes in different galaxies, but there are also different offsets, particularly 
in post-starburst galaxies \citep[see][]{French15}.  And, if no one such relation holds for all galaxies, then the 
S-K relation does not represent a universal physical law of star formation \citep[see][]{Lada13}.   Perhaps the S-K relation for a 
disk galaxy is a measure of how the inverse depletion time of the molecular gas varies radially in that disk galaxy
(e.g., inside-out or outside-in star formation), which, in turn, is affected by the many properties of, and processes in, the disk of that 
particular galaxy \citep[see, for example][]{Meidt13, Hughes13, Schinnerer13, Colombo14}.

\section{Summary and Conclusions}\label{sumcon}

The spiral galaxies \mfif\ and \meig\ were observed with the bolometer array {\it AzTEC\/} on the {\it JCMT\/} in the 
1.1$\,$mm continuum at 20$''$ spatial resolution.  The extended emission, including the interarm emission and exponential 
disks, was faithfully recovered in the final maps out to galactocentric radii of more than 12$\,$kpc for both galaxies. 
This was verified by simulations that show that only small corrections are necessary for the \mfif\ image and negligible 
corrections for that of \meig.   The 1.1$\,$mm-continuum fluxes are $5.6\pm 0.7$ and $9.9\pm 1.4\, Jy$ for \mfif\ and
\meig, respectively.  The uncertainties are largely due to that of the calibration. 

These images were combined with the 160$\um$ image of {\it Spitzer\/} to obtain dust temperatures and column densities.
This approach was adopted, rather than using dust models, for two reasons.  One reason was to have an independent test 
of the models \citep[see][]{Groves15}.  The other reason is that multi-wavelength far-IR data at the long 
wavelengths (i.e. $\lambda\gsim 100\um$) at which the bulk of the dust mass radiates is not available for the outer 
disks of \mfif\ and \meig, other than the 160$\um$ data of {\it Spitzer\/}. 

Another model-independent approach was to constrain, at least roughly, the $\taun$ at 1.1$\,$mm 
by the observations, rather than simply adopting a dust mass absorption coefficient, $\knup$.  
Gas column densities were estimated by calibrating against HI-dominant positions to estimate the dust optical depth to
gas column density ratio $\taun$. 
The method of calibrating against the HI-dominant positions was improved by crudely estimating
the effects of the CO-dark gas (see below).  
Neither galaxy has a strong radial variation in the gas surface density beyond 
galactocentric radii of about 3$\,$kpc.   Out to a galactocentric radius of 14$\,$kpc, the best estimate of the mass of gas 
in \mfif\ is $9.4\times 10^9\, M_\odot$.  Out to 12$\,$kpc in \meig, this best estimate is $7.2\times 10^9\, M_\odot$.
(See adopted distances in Table~\ref{tab1}.)

Pre-existing maps of CO~$\Jone$ permitted the creation of maps of the $\NH/I(CO)$ or X-factor for both \mfif\ and \meig\
out to galactocentric radii of 6-8$\,$kpc.  Both galaxies have X-factor values that are higher in the {\it inter}arm than
in the arms by a factor of $\sim 1.5$-2.  In the central few kiloparsecs 
of \mfif, this interarm/arm ratio rises to $\sim 3$.  
In \mfif, there is no significant radial variation of $\Xf$.  In \meig, however, there is evidence at the 
many-$\sigma$ level of radial variation of the X-factor by factors of 2 to 3, where the central 2$\,$kpc radius has a roughly
flat X-factor which declines to a minimum at about 4$\,$kpc.  Within galactocentric radii $\leq 6$-8$\,$kpc, 
the spatially averaged X-factor is about 1$\Xtw$. 

Beyond the outer radius of the X-factor map for each galaxy, comparison
of the radial profile of the gas surface density derived from the continuum with those of the surface brightnesses of CO and HI
permits estimates of lower limits of the X-factor that reach one to two orders of magnitude higher than the inner disk values.
This suggests the existence of CO-dark molecular gas in the outer disks of \mfif\ and \meig, although alternative explanations
are only partly ruled out.  Nevertheless, these alternatives do not entirely account for the high X-factor values in the outer disk.

A two-dimensional Fourier analysis of the spiral structure at 1.1$\,$mm and at visible (or near visible) wavelengths revealed
that the spiral structure in red light and that in the 1.1$\,$mm continuum and were the same in \mfif.  For \meig, 
the spiral structure in $I$-band compared with that in the 1.1$\,$mm continuum showed that the bar's effect
in \meig\ is conspicuous in $I$-band and not at 1.1$\,$mm.  These results suggest that the spiral density wave in \mfif\ is 
influencing the interstellar medium and stars similarly, while the bar potential in \meig\ has a different influence on the 
interstellar medium from that on the stars. 

Log-log plots of the star formation rate surface densities against those of the gas traced by spectral lines (i.e., of HI and CO) 
had slopes of $\sim 1.5$ whether total gas or just molecular gas surface density.  For the plots with gas surface densities 
traced by the continuum emission, the slopes were $\sim 2.5$ whether
total gas (using continuum only) or just molecular (using continuum with HI subtracted) gas surface densities.  These plots, especially
with the continuum tracers, show a threshold gas surface density at which the SFR rises by two or more orders of magnitude.  The 
existence of this threshold gas surface density is insensitive to within a factor of $\sim$3 for the adopted $\taun$.   The value of this 
threshold density is $\sim 15\mspc$. 
This threshold is somewhat less conspicuous in the spectral line tracers than for the continuum tracers.

The fitted slopes suggest that the depletion of the molecular gas occurs first at small galactocentric radii and then at increasing 
radii in an inside-out galactic evolution.  This is seen more clearly in maps of the ratio of the surface densities of the SFR to 
that of the molecular gas.  
For both these galaxies and the continuum tracer, the 
molecular gas depletion time in the centres is about 1$\,$Gyr, rising at radii of 6-8$\,$kpc to around 10-20$\,$Gyr.  Further out, 
the depletion times rise by one or two orders of magnitude.  The spectral line tracer, i.e. CO~$\Jone$, suggests molecular gas 
depletion times in the outer disks that are appreciably less than 10-20$\,$Gyr.  

The images of the inverse depletion time show signs of spiral structure.  Superficially, this suggests that spiral arms
effect the SFR beyond just heightening the gas surface density.  However, correcting for the X-factor spatial variation and
for the spatial variation of the $\taun$ removes or nearly removes such spiral structure.  This apparently confirms the result 
of \citep{Foyle10} that the arms merely heighten the SFR in the same proportion as they heighten the gas surface density. 
Greater spatial resolution is required for confirming this result. 

In the future, we need deeper mapping of molecular tracers in the outer disks of these spiral galaxies.  As well, 
a better method of calibrating the $\taun$ at millimetre wavelengths is needed.  So far, the method employed in the current 
work is functional, but only crudely.  Either a new method or refinement of the method described here is necessary. 

\bigskip
\noindent  Consultations with Rich Rand, Divakara Mayya, Daniel Rosa are greatly appreciated.  We also
thank the anonymous referee, whose comments noticeably improved the manuscript.

\appendix

\section{Flux Determination}\label{appflux}

The outer edge to the 1.1$\,$mm continuum emission of \mfif\ lies at a galactocentric radius
of about 14$\,$kpc and for \meig\ it is about 12$\,$kpc.   
The derived fluxes are $5.6\pm 0.7\, Jy$
for \mfif\ and $9.9\pm 1.4\,Jy$ for \meig\ to those outer edges. 
The uncertainties include the calibration uncertainty
of 13\% and the uncertainties due to the somewhat arbitrary choices for outer boundaries.  The 
simulations were used to check for systematics in the flux determination, thereby providing correction
factors for the derived fluxes.  For \mfif, this correction factor was 1.5 for \mfif\ and 0.9 for \meig.
These corrections were included in the previously quoted flux values.  

\section{Determination of Surface Densities and Masses}\label{appsurf}

The 1.1$\,$mm continuum surface-brightness maps of Figures~\ref{fig1} and \ref{fig2}  were ratioed 
with the {\it Spitzer/MIPs\/} 160$\um$, effectively 155.9$\um$, maps 
of \mfif\ and \meig\ \citep{Dale09} to produce temperature maps under the assumption that the dust 
emissivity index, $\beta$, 
was 2.0.  As stated in the Introduction, this was
the most appropriate value for $\beta$ as found from the observations of $\meig$ by \citet{Foyle12}.   
{\it In practice, the adopted value for $\beta$ will have only a small effect 
on the derived gas column densities, provided that those derived column densities are calibrated against observed
gas column densities.\/}  By using those positions where atomic gas dominated the column densities to calibrate the
$\taun$ for the continuum-derived column densities, it was found that the column densities derived from the continuum 
data for $\beta=1.5$ to 2.5 did not deviate by more than 30\% from those for $\beta=2.0$. Consequently, observations at
only two wavelengths are necessary for a reasonable approximation of the gas surface density map.  
  
The $\taun$ derived from the HI comparisons are 2.2-$2.3\times 10^{-26}\, cm^2$ for \mfif\ and \meig, respectively.
These values lead to gas mass estimates that appear to 
be too low (see Sections~\ref{Xfac}, \ref{xfsv}, and below in this appendix for details).  
\citet{Foyle12} simply adopted the dust mass-absorption coefficient of 
\citet{Li01} and \citet{Draine03}, which is a factor of $\sim 6$ smaller than the values determined here. 
\citet{Foyle12} also determined a higher dust-to-gas mass ratio than that adopted here.  This is equivalent to adopting a 
$\taun$ that is factors of 5.0-5.3 smaller than that derived from the HI comparisons. 
Because we adopt a constant dust-to-gas ratio,  the two $\taun$ cases 
are referred to as the high-$\knu$ (from HI comparisons) and low-$\knu$ \citep{Li01, Draine03} cases.  
Given the low masses that result from the high-$\knu$ case, the HI comparisons must be appropriately modified.
The HI comparisons were repeated after removing upper outliers that are apparently unreliable due to possible 
undetected gas.  This results in an intermediate 
case, where a $\taun$ that is a factor of two lower than that from the simple atomic gas calibration method
is found.  Unless otherwise
stated, all masses and surface densities quoted are for this intermediate-$\taun$ (most realistic) case. 
This approach yields a roughly 50\% uncertainty in column density and mass estimates, using $\taun = 1.1\times 10^{-26}\, cm^2$. 

The surface brightness at frequency $\nu$, $\Inu(\Td)$, is related to the dust-derived gas column density, $\Nd$, by
\begin{equation}
\hfill\Inu = \mu m_H x_d\, \knu \Nd \Bnu(\Td)\qquad ,\hfill
\label{eq1}
\end{equation}  
where $\mu$ is the mean atomic weight per hydrogen atom, $m_H$ is the mass of a hydrogen atom, $x_d$ is 
the dust-to-gas mass ratio, and $\knu$ is the dust
mass absorption coefficient.   The mass-absorption coefficient varies with frequency as follows:
\begin{equation}
\hfill\knu = \kno\, {\left(\nu\over\nu_o\right)}^\beta\qquad ,\hfill
\label{eq2}
\end{equation}
in which $\kno$ is dust mass-absorption coefficient at a reference frequency $\nu_o$.   The dust
temperature, $\Td$, is of course estimated by evaluating expression~(\ref{eq1}) at frequency $\nu_1$, corresponding
to the wavelength of 1.1$\,$mm for the {\it JCMT/AzTEC\/} observations, and again at frequency $\nu_2$, corresponding 
to the wavelength of 155.9$\um$ for the {\it Spitzer/MIPS\/} observations, and taking the ratio of the two.

As well as using the maps at the two wavelengths (i.e. the {\it JCMT/AzTEC\/} at 1.1$\,$mm and the
{\it Spitzer/MIPS\/} map at 155.9$\um$),
there are many additional details involved in estimating the column densities, $\Nd$.  These include estimating 
the value of $\taun$ (or the product of $x_d$ and $\knu$), convolving the {\it AzTEC\/} 1.1$\,$mm map to the 
resolution of the {\it Spitzer/MIPS\/} 155.9$\um$ map, estimating the noise levels in the maps, characterizing
any systematic effects introduced into the $\Td$ and $\Nd$ maps due to the observations and processing of the 
{\it AzTEC\/} data, and any relevant colour corrections to the continuum data.  In particular, a
constant offset correction of $5.5\times 10^{20} H$-$nuclei\cdot cm^{-2}$ must be added to the column densities for 
\mfif\ as dictated by the simulations to correct for inadequately recovering the large-scale emission (but was not 
applied to the figures).
\subsection{The Reliability of Gas Surface Densities and Masses}\label{appgsdm}

There are a couple issues that should be addressed in discussing the reliability of gas surface densities and masses as inferred from dust 
emission in the current work.  One issue is to what level calibration differences between that of {\it SPITZER/MIPS\/} and that of 
{\it HERSCHEL/PACS\/} for their 160$\um$ data would affect the results and conclusions of the current work \citep[see][]{Aniano12}.  
Another issue is the validity of the simple approach used here instead of using a more detailed dust model \citep[e.g.,][]{Li01}.  
As for the former issue, \citet{Aniano12} compared the 160$\um$ data for both instruments and characterized their systematic differences.
Depending on the calibration scheme used, the {\it HERSCHEL/PACS\/} 160$\um$ intensities were either 1.6\% or 25\% systematically higher 
than those for {\it SPITZER/MIPS\/}.  
Scaling up the 160$\um$ intensities used in the current work by 25\% would increase the computed
dust temperatures, but by differing amounts depending on initial computed temperature.   Most of the those temperatures were determined
to be between 10$\,$K and 23$\,$K.  After the hypothetical correction, those would be 10.3$\,$K and 25$\,$K.  If there were no calibration
of the column densities against other data, those corrected temperatures would require the column densities to be corrected downwards
by 5\% and 11\%, respectively.   The warmer positions in both galaxies tend to have higher column densities.  So the higher column
densities would be corrected downwards by more than the lower column densities, on average.  This reduces the dynamic range of the 
determined column densities by roughly 6\%.  The overall scaling of those column densities would remain unchanged because of the 
calibration against HI column densities.   So, such a correction, were it necessary, would not appreciably change the current
results. 

As for the second issue, the models of \citet{Li01} combined with data at mid-IR and far-IR wavelengths
can provide estimates of a number of parameters.
For the current work, however, there
are two reasons why such an approach was not used.   One reason is that, as mentioned in the introduction, the {\it AzTEC\/} 1.1$\,$mm map has more
spatial coverage than the {\it HERSCHEL\/} data for both \mfif\ and \meig. {\it SPITZER\/} does have sufficient spatial coverage for comparison
with the {\it AzTEC\/} data, but does not have the long-wavelength data that probes the bulk of the dust mass, other than at 160$\um$.  So, to
consistently treat each entire galaxy, the {\it AzTEC\/} 1.1$\,$mm data were combined with the {\it SPITZER\/} 160$\um$ data for both
\mfif\ and \meig.  Another reason that the more model-dependent approach was not used is that we need to have model-independent tests that can 
confirm or refute the validity of the models, as stated recently by \citet{Groves15}.  The current work, for example, finds a 
dust opacity to gas column density ratio $\taun$ at 1.1$\,$mm that is about double that inferred from \citet{Li01}.   But temporarily 
adopting the $\taun$ of \citet{Li01} and adopting the appropriate distance yields nearly identical masses for both the more
sophistocated approach \citep[i.e., in][]{Cooper12} and the simpler approach (i.e., the current work) for the NGC$\,$5194 field.  This strongly
suggests that the simpler approach of the current work is reasonable at recovering some basic dust properties.  

The dust optical depth to gas column density ratio, $\taun$, and especially 
the dust mass absorption coeffient, $\knu$, can be quite uncertain, ranging from $\knup=0.06$ to $1.4\cmg$. \citep[see the following for examples 
of disparate values:][]{Eales10, Li01, Draine03, James02, Dunne11, Oss94}.   Accordingly, attempting to measure $\taun$ observationally 
could narrow the uncertainty.  This is most easily observed in gas dominated by atomic hydrogen because the gas column densities are easy to
determine.  Another approach was followed by \citet{Eales10} for observations of the galaxies M$\,$99 and M$\,$100 in which they plotted the 
star formation rate surface density, $\Sfr$, against gas surface density and against that for dust, adjusting the $\knu$ until both surface densities
agree at large $\Sfr$.  This resulted in a $\knu$ value corresponding to the unusually low $\knup=0.06\cmg$ mentioned above.  

The large $\taun$ initially derived in the current work was from comparing the observed optical depths at 1.1$\,$mm 
with the observed gas column densities in regions dominated by 
atomic gas.   The \citet{Parade11a,Parade11b} observed $\taun=5.2\times 10^{-26} cm^2$ at 857$\,$GHz and 
$\taun=1.1\times 10^{-25} cm^2$ at 250$\um$ in the dust associated with the HI gas within the Galaxy. 
The \citet{Parade11b} also observed $\taun = 2.3\times 10^{-25} cm^2$ at 250$\um$ for the dust associated with molecular 
gas, which is about double that observed for dust associated with atomic gas, remembering that 
the $N(H)$ is the column density of hydrogen {\it nuclei\/}.  At a wavelength of 1.1$\,$mm, these numbers 
correspond to $6.6\times 10^{-27} cm^2$ and $7.9\times 10^{-27} cm^2$ for dust associated with HI 
and $\taun = 1.6\times 10^{-26} cm^2$ for that associated with H$_2$ \citep[using a spectral 
emissivity index of $\beta=1.8$ for the Galaxy, e.g.,][]{Parade11a, Parade11b}.  The current work initially finds 
$\taun=2.2\times 10^{-26} cm^2$ for \mfif\ and $2.3\times 10^{-26} cm^2$ for \meig\ for dust along lines of 
sight with gas column densities dominated by $HI$.
This is more than double the values observed in our Galaxy and are even 40\% larger than the dust associated with $H_2$ 
in our Galaxy.  This large $\taun$ is suspect and requires a re-evaluation of the calibration of $\taun$ at 1.1$\,$mm.


One explanation for the observed large $\taun$ would be CO-dark gas.  
The presence of such gas can be crudely tested.  This is done by starting with dust-derived gas column densities, $\Nd$, that are only roughly 
calibrated -- i.e., to within a factor of a few.  The difference $\Nd - N(HI)$ then provides a rough measure of molecular gas column
density.  The ratio $(\Nd - N(HI))/(2\,\Xf I(CO))$ (where the $\Xf$ is the average used for each galaxy) is a crude measure of the 
amount of dark gas --- a {\it dark\/} ratio.   Then $\taun$
can be recomputed using only those positions with dark ratios less than some upper limit.   
Choosing a dark ratio upper limit as 
high as 100 reduces the derived $\taun$ by a few percent, while choosing unity yields a $\taun$ that is a factor of about 4 lower.
The latter is risky given that 
this could unnecessarily exclude too many positions with very little dark gas.  
Accordingly, a more conservative correction 
factor would be around 2.  

Reducing the $\taun$ value by this factor scales up the column densities by the same factor and the X-factor 
scales in proportion to $\Nd - N(HI)$, or up by a factor of $\sim$3 for both \mfif\ and \meig.  The best estimates of these
values are hence $\taun\simeq 1\times 10^{-26} cm^2$, corresponding to $\knup\simeq 0.7\cmg$ for $\mu x_d = 0.01$ for both \mfif\ and \meig.
These are well within a factor of 2 of the values found by the \citet{Parade11a,Parade11b} for our Galaxy.  The best estimates for the X-factors
for \mfif\ and \meig\ are around 1$\Xtw$.  Table~\ref{tab3} displays the best estimated values for the gas masses and the X-factors.  That
table shows that roughly half of the total mass of the gas is unaccounted for using spectral lines.  This is discussed further in 
Section~\ref{xfsv}.

Possible spatial variations in $\taun$ should also be considered.  
Both \mfif\ and \meig\ show
{\it de}creases in $\taun$ by about 15-20\% in going from galactocentric radii of 1 or 2$\,$kpc to about 6 or 7$\,$kpc.
Correcting for the CO-dark gas changes this result.  The problem is that the numbers of points
used in the comparisons becomes very small (i.e. $\sim 10$) when the upper limit to the dark ratio is reduced.  Such small numbers do not 
give reliable estimates of $\taun$.  So the radial spatial variation of $\taun$ is uncertain, although there is no clear evidence that it has
a strong radial variation.  In going from interarm positions to spiral arm positions, the $\taun$ in \mfif\ decreases by 18\%.  In \meig, this
decrease is 8\%.  Again, correcting for the CO-dark gas changes these results to 25\% for \mfif\ and an 11\% {\it in}crease for \meig, although 
the number of points available for the estimation is not really sufficient.


\section{Determination of the X-Factor Maps}\label{appx}

The X-factor, $\Xf$, can be straightforwardly computed from
\begin{equation}
\hfill\Xf = {\Nd - N(HI)\over 2\, I(CO)}\qquad . \hfill
\label{eq5}
\end{equation}
Given that $\Nd$, $N(HI)$, and $I(CO)$ are all maps, $\Xf$ is also a map, but
not a very extended map because of the mounting uncertainties that
result from the numerous computations with pre-existing data required to reach $\Xf$.  
Both the \mfif\ and \meig\ $\Xf$ maps do not extend
beyond about 7$\,$kpc from the centre of each map, mostly because the CO maps do not extend further
than that.  Figures~\ref{fig13}, \ref{fig13a}, \ref{fig14}, and \ref{fig14a} show the X-factor maps and 
radial profiles for \mfif\ and \meig.  The mean values of the X-factor listed in Table~\ref{tab5}, 
which we call $\Xfm$, are the $1/\sigma(\Xf)^2$-weighted means of each $\Xf$ map and only really 
applies to the central 7$\,$kpc radius in each galaxy. 

Given that the CO~$\Jone$ emission falls to the noise level at a radius of about 7 to 8$\,$kpc, we are unable to compute
$\Xf$ beyond that in the outer disks.  Nevertheless, we can use the radial curves of Figures~\ref{fig5a} and \ref{fig6a} 
to infer rough lower limits on $\Xf$ in the outer disks.  The radial variation of $\Xf$ with respect to the mean value
in the inner disk, $\Xfm$, is a simple variation on expression~(\ref{eq5}):
\begin{equation}
\hfill{\Xf\over\Xfm} = {\Nd - N(HI)\over 2\,\Nxh}\qquad , \hfill
\label{eq6}
\end{equation}
where $\Nxh$ is the molecular gas column density estimated from a spatially constant X-factor: $\Nxh\equiv\Xfm I(CO)$.
This measure of the column density is very noisy in these observations of the outer disks and a useful proxy is necessary
for inferring some crude lower limit on $\Xf/\Xfm$ for radii beyond 8$\,$kpc.  
The proxy adopted for $2\Nxh$ was a pessimistic estimate of the 3-sigma uncertainties of the gas column density,
i.e., $3[\sigma(N(HI))+2\sigma(\Nxh)]$. This gives the 
most conservative (i.e. least extreme) estimates for $\Xf/\Xfm$ beyond a radius of 8$\,$kpc.  For \mfif, even in the 
high-$\knu$ (low-$\knu$) case, the above proxy yields lower limits to $\Xf$ of roughly $1$ to $30\Xtw$ 
($7$ to $10^3\Xtw$) depending on the radius in the outer disk.  For \meig, even in the high-$\knu$ (low-$\knu$) case, 
these rough lower limits to $\Xf$ are $0.3$ to $20\Xtw$ ($2$ to $600\Xtw$).   

\section{Considerations of the Comparison between SFR and Gas Surface Densities}\label{appsfrgas}

The logarithm of
the SFR surface density is plotted against that of the gas surface density for different tracers of that gas surface density.  
The plots of the surface densities of SFR versus gas are displayed in Figure~\ref{fig18}.
The uncertainty of each fitted slope is the formal error of the fit and was 
scaled by the square-root of the reduced chi-square in order to give a more conservative and more realistic estimate of this 
formal error.  One important consideration in these 
fits is that the spatial resolution of the \mfif\ data is limited to 38$''$,  because of the {\it Spitzer\/} 160$\um$ resolution.  
For \meig\ this resolution is 55$''$ resolution due to the CO observations.  Given that each point in these plots represents a 
single pixel, the effective number of independent points in each plot is the number of pixels divided by 20 for \mfif\ and by 
55 for \meig.   Nevertheless, the linear resolution on each galaxy is about the same, with 1.5$\,$kpc for \mfif\ and 1.2$\,$kpc for \meig.

\section{Considerations for Creating Star Formation Rate Surface Density Maps}\label{appsfrsurf}

The SFR surface density maps were created from images in the H$\alpha$ line and in the 24$\um$ continuum.  According to \citet{Calzetti07}, 
the SFR, $\SFR$, within a source is given by 
\begin{equation}
\hfill \SFR = a[L(H\alpha) + b\,\nu L_\nu(24\um)]\qquad .\hfill
\label{eqb1}
\end{equation}
$L(H\alpha)$ is the integrated luminosity of the H$\alpha$ line in $\es$; $\nu$ is the frequency corresponding to the
wavelength of 24$\um$; $L_\nu(24\um)$ is the luminosity of the 
24$\um$ continuum per unit frequency bandwidth in  $\esh$; $a = 5.3\times 10^{-42}\, M_\odot\cdot yr^{-1}
\cdot(\es)^{-1}$; $b=0.031$; $\SFR$ is in units of $M_\odot\cdot yr^{-1}$.  If $\Sfr$ is the SFR surface density, then
$\Sfr = \SFR/(\Omega_s D^2)$, where $\Omega_s$ is the source solid angle and D is the source distance.  Using $L=4\pi D^2 F$, where
$F$ is the source flux, and also using $I = F/\Omega_s$, with $I$ as the source surface brightness, results in 
\begin{equation}
\hfill \Sfr = 4\pi a[I(H\alpha) + b\, c_\nu I_\nu(24\um)]\qquad .\hfill
\label{eqb2}
\end{equation}
For $\Sfr$ in units of $\msyrkp$, $I(H\alpha)$ in units of $\escs$, and $I_\nu(24\um)$ in 
units of $\mjsr$, we have $a=5.1\times 10^1\msyrkp\cdot (\escs)^{-1}$ and $c_\nu =
1.249\times 10^{-4}\, \escs\cdot (\mjsr)^{-1}$.  The numerical value of $b$ remains
unchanged.

Specifically for \mfif\ and \meig, the SFR surface density maps were created by applying expression~(\ref{eqb2}) to the 
H$\alpha$ \citep[i.e., the line-integrated images of][]{Kennicutt03, Blasco10} and the Spitzer 24$\um$ images \citep{Dale09} 
of both galaxies.  The images were convolved to the appropriate resolution and rebinned to 9$''$ pixels. The $I(H\alpha)$ image 
was converted to units of $\mu\escs$.  The final spatial resolution of the SFR surface density map for \mfif\ is 38$''$,
which is that of the {\it Spitzer\/} 160$\um$ image.   For \meig\ this is 55$''$, which is that of the CO~$\Jone$
map. The negative tails of the histograms of pixel values provided estimates of the noise levels. 


\section{Relationship between the X-Factor and the Star Formation Rate Surface Density}\label{appxfsfr}

The physical models of \citet{Clark15} explore the possible dependence of the X-factor on the
the star formation rate. In their Section~5.1, their results can be parameterized in the form 
\begin{equation}
\hfill \Xf = k_{xs} \Sfr^\gamma \qquad ,\hfill
\label{eqxs1}
\end{equation}
where they find $\gamma\simeq 0.5$.  With the results of the fits illustrated in Figure~\ref{fig18},
we can test this result as well as specify the parameter $k_{xs}$. 

We repeat the derivation in Section 5.1 of \citet{Clark15} in more detail and adopt their nomenclature
where applicable.  If CO (1-0) is used to estimate the H$_2$ surface density, $\Sigma_{mol}$, then we
let $\Sigma_{mol}(CO)$ represent this surface density as estimated by CO.   Then
\begin{equation}
\Sfr = k_{sco} \left[\Sigma_{mol}(CO)\right]^{N_{obs}} = 
k_{sco} \left[\Xfm I(CO)\right]^{N_{obs}}\ ,\hfill
\label{eqxs2}
\end{equation}
where $N_{obs}$ is the power-law index that is observed when using CO-derived molecular gas surface
densities and $\Xfm$ is the mean X-factor value chosen for the galaxy observed.\footnote{Notice that
the value chosen for $\Xfm$ is unimportant because it is the $k_{sco}\Xfm^{N_{obs}}$ combination that
matters.} Using a tracer that supposedly gives the ``true'' $\Sigma_{mol}$, would have a similar relation:
\begin{equation}
\Sfr = k_{sh} \left[\Sigma_{mol}\right]^{N_{act}} = 
k_{sco} \left[\Xf I(CO)\right]^{N_{act}}\qquad ,\hfill
\label{eqxs3}
\end{equation}
where $N_{act}$ is the actual correct power-law index and $\Xf$ is the correct point-by-point
value of the X-factor.  Combining expressions~(\ref{eqxs2}) and (\ref{eqxs3}) so as to eliminate
$I(CO)$ and comparing with (\ref{eqxs1}) yields
\begin{equation}
k_{xs} = \Xfm {k_{sco}^{1/N_{obs}}\over k_{sh}^{1/N_{act}}}\qquad ,\hfill
\label{eqxs4}
\end{equation}
and also,
\begin{equation}
\gamma = {1\over N_{act}} - {1\over N_{obs}}\qquad .\hfill
\label{eqxs5}
\end{equation}

In the context of the current work, we assume that the dust tracers of the molecular
gas give $N_{mol}$ (i.e., \hbox{$\Nd - N(HI)$}).  So the fits in the second panels of 
Figure~\ref{fig18} yield the parameters $k_{sh}$ and $N_{act}$ for \mfif\ and \meig.
The fits in the fourth panels provide $k_{sco}$ and $N_{obs}$.
The uncertainties of these quantities are likely dominated by systematics.  This
means that $k_{xs}$ has a roughly 50\% uncertainty.  The effects of systematics on
$\gamma$ are difficult to estimate.  The formal uncertainties of $\gamma$ using the 
current data are 3-4\%. 

With the above in mind, expressions~(\ref{eqxs4}) and (\ref{eqxs5}) applied to the current
data yield $k_{xs} = 0.18$ and $\gamma = -0.38$ for \mfif\ and 0.79 and $-0.25$ for \meig. 
Note that these values for $k_{xs}$ give $\Xf$ in $\Xtw$ units. 

\label{lastpage}

\end{document}